\documentclass[aps,prd,superscriptaddress,floatfix,reprint]{revtex4-1}
\pdfoutput=1
\usepackage{graphicx}
\graphicspath{{plots/}}
\usepackage{amsmath}
\usepackage{amssymb}
\usepackage{bbm}
\usepackage{hyperref}

\DeclareMathOperator\sign{sign}
\DeclareMathOperator\diag{diag}
\DeclareMathOperator\Tr{Tr}
\DeclareMathOperator\re{Re}

\renewcommand\epsilon\varepsilon
\renewcommand\phi\varphi
\newcommand\q{q}
\renewcommand\d{d}
\newcommand\zm{\text{zm}}
\newcommand\nzm{\text{nzm}}
\newcommand\U{\text{U}}
\newcommand\SU{\text{SU}}
\newcommand\1{\mathbbm{1}}
\newcommand\x{x}
\newcommand\m{u}
\newcommand\hu{\hat u}
\newcommand\M{\vec u}
\newcommand\Kuphi[1]{K_{#1}(2|\m\sin\tfrac\phi2|)}
\newcommand\Khuphi[1]{K_{#1}(2|\hu\sin\tfrac\phi2|)}

\def\Xint#1{\mathchoice
   {\XXint\displaystyle\textstyle{#1}}%
   {\XXint\textstyle\scriptstyle{#1}}%
   {\XXint\scriptstyle\scriptscriptstyle{#1}}%
   {\XXint\scriptscriptstyle\scriptscriptstyle{#1}}%
   \!\int}
\def\XXint#1#2#3{{\setbox0=\hbox{$#1{#2#3}{\int}$}
     \vcenter{\hbox{$#2#3$}}\kern-.5\wd0}}
\def\dashint{\,\Xint-}

\begin{document}

\title{\boldmath Dirac spectrum and chiral condensate for QCD at fixed $\theta$-angle}

\author{M.~Kieburg} 
\affiliation{Department of Physics, University of Bielefeld, 33501 Bielefeld, Germany}
\email{mkieburg@physik.uni-bielefeld.de}

\author{J.J.M.~Verbaarschot}  
\affiliation{Department of Physics and Astronomy, SUNY, Stony Brook, New York 11794, USA}
\email{jacobus.verbaarschot@stonybrook.edu}

\author{T.~Wettig}  
\affiliation{Department of Physics, University of Regensburg, 93040 Regensburg, Germany}
\email{tilo.wettig@ur.de}

\date{\today}

\begin{abstract}
  We analyze the mass dependence of the chiral condensate for QCD at
  nonzero $\theta$-angle and find that in general the discontinuity of the
  chiral condensate is not on the support of the Dirac spectrum. To
  understand this behavior we decompose the spectral density and the
  chiral condensate into contributions from the zero modes, the
  quenched part, and a remainder which is sensitive to the fermion
  determinant and is referred to as the dynamical part. We obtain
  general formulas for the contributions of the zero
  modes. Expressions for the quenched part, valid for an arbitrary
  number of flavors, and for the dynamical part, valid for one and two
  flavors, are derived in the microscopic domain of QCD. We find that
  at nonzero $\theta$-angle the quenched and dynamical part of the Dirac
  spectral density are strongly oscillating with an amplitude that
  increases exponentially with the volume $V$ and a period of order of
  $1/V$. The quenched part of the chiral condensate becomes
  exponentially large at $\theta\ne0$, but this divergence is canceled by the
  contribution from the zero modes. The oscillatory behavior of the
  dynamical part of the density is essential for moving the
  discontinuity of the chiral condensate away from the support of the
  Dirac spectrum. As important by-products of this work we obtain
  analytical expressions for the microscopic spectral density of the
  Dirac operator at nonzero $\theta$-angle for both one- and two-flavor QCD
  with nonzero quark masses.
\end{abstract}

\maketitle

\allowdisplaybreaks[3]

\section{Introduction}

Topology in the form of instantons and dyons is an important
ingredient of the QCD vacuum
\cite{tHooft:1976snw,Belavin:1975fg,Gross:1980br}. The main reason is
that the Dirac operator for field configurations with nonzero
topological charge has a generic number of exact zero modes, which
induce the chiral condensate for massless quarks. At nonzero quark
mass, the total number of instantons and anti-instantons is even more
important. This number scales with the four-dimensional space-time
volume $V$, unlike the topological charge, which scales as $\sqrt V
$. If instantons and anti-instantons are not strongly overlapping they
give rise to near-zero modes \cite{Schafer:1996wv} which determine the
value of the chiral condensate. Given the importance of topology for
the QCD partition function, it is puzzling that the conjugate
parameter, the so-called $\theta$-angle, is consistent with zero according
to all available experimental evidence. Nevertheless, theories with
nonzero $\theta$-angle have received a great deal of attention both as
theories beyond the standard model as well as from a purely
theoretical perspective
\cite{Leutwyler:1992yt,Azcoiti:2003ai,Kanazawa:2015uvg,Cai:2016eot,Gaiotto:2017tne,Tanizaki:2017bam}.

In \cite{Verbaarschot:2014upa,Verbaarschot:2014qka} we have resolved
an interesting apparent puzzle for one-flavor QCD at zero
$\theta$-angle: the chiral condensate should be independent of the sign of
the quark mass $m$, but the condensate expressed in terms of the Dirac
eigenvalues appears to be an odd function of $m$. The point is that
this function still needs to be averaged over gauge fields, and this
average depends on the quark mass through the fermion determinant in
such a way that the chiral condensate eventually becomes an even
function of $m$.  This resolution is reviewed in
Sec.~\ref{sec:review}: at negative quark mass, the statistical weight
in the average over gauge fields becomes negative, which leads to
exponentially large oscillations that can move the discontinuity of
the chiral condensate away from the support of the Dirac spectrum and
could be shown to yield a mass-independent condensate.  In the present
paper we extend the work of
\cite{Verbaarschot:2014upa,Verbaarschot:2014qka} to arbitrary
$\theta$-angle and to more than one flavor, as already sketched in
\cite{Kieburg:2017kex}. Let us emphasize that for nonzero $\theta$ the
statistical weight is not only negative but becomes complex.

To be able to obtain explicit results, most of our calculations are
performed in the $\epsilon$-domain of QCD (also called microscopic
domain). In this domain the Compton wavelength of the pion is much
larger than the size of the box so that the kinetic term of the chiral
Lagrangian can be ignored and only the mass term remains. This chiral
Lagrangian can also be obtained from a random matrix theory with the
same global symmetries as QCD
\cite{Shuryak:1992pi,Verbaarschot:1994qf}, which makes it possible to
find analytical expressions for the spectral density of the Dirac
operator using powerful random matrix techniques.  Another benefit of
working in the $\epsilon$-domain is that the chiral condensate can be
obtained from the spectral density without any additional
regularization. This is important because the spectral density of the
Dirac operator is renormalization-group and gauge invariant, while the
chiral condensate is only gauge invariant.

The structure of this paper is as follows. In Sec.~\ref{sec:review} we
review the mass dependence of the chiral condensate for one- and
two-flavor QCD. The sign problem for QCD at nonzero $\theta$-angle is
discussed in Sec.~\ref{sec:sign}.  A decomposition of the spectral
density and the chiral condensate is introduced in
Sec. \ref{sec:decom}. In Sec.~\ref{section-4} we derive general
analytical expressions for the contributions of the zero modes and of
the quenched part of the Dirac spectrum to the chiral condensate and
show that each of them increases exponentially with the volume at
nonzero $\theta$-angle, but that their sum remains finite. The one-flavor
case is worked out in detail in Sec.~\ref{sectio-5}, where we also
obtain an expression for the spectral density of the Dirac operator at
fixed $\theta$-angle. The two-flavor case is discussed in
Sec.~\ref{section-6}, where we derive analytical expressions for the
spectral density and the chiral condensate at fixed
$\theta$-angle. Concluding remarks are made in Sec.~\ref{sec:conclusio},
and technical details are given in several appendices.  In particular,
in App.~\ref{sec:resum} we obtain identities for sums of products of
three and four Bessel functions which, as far as we know, are new.

Some notes on notation: on the macroscopic scale, the Dirac
eigenvalues and the quark masses are denoted by $\lambda$ and $m$, while on
the microscopic scale they are denoted by $\x$ and $\m$,
respectively. For functions of these variables, such as the partition
function $Z$, the spectral density $\rho$ or the chiral condensate
$\Sigma$, we use the same symbol on the macroscopic and the microscopic
scale to simplify the notation.  The corresponding functions are of
course different, but it should be clear from the arguments of the
function what is meant in every case. Also, when we give results for
partition functions we drop irrelevant normalization factors.

\section{Review of known results}
\label{sec:review}

We consider QCD with $N_f$ quark flavors and quark mass matrix
$M=\diag(m_1,\ldots,m_{N_f})$, which we allow to be complex for the time
being. The mass matrix appears in the QCD Lagrangian in the form
$\bar\psi_R M\psi_L+\bar\psi_L M^\dagger\psi_R$, where $\psi_{R/L}$ denotes quark fields of
definite chirality.  For a given gauge-field configuration with
topological charge $\nu$, the fermion determinant is
\begin{align}  
  \prod_{\lambda_n>0}\det(\lambda_n^2+MM^\dagger)
  \times
  \begin{cases}
    \det^\nu M, & \nu\ge0\,,\\
    \det^{-\nu} M^\dagger, & \nu<0\,,
  \end{cases}
\end{align}
where the $\lambda_n$ are the eigenvalues of the Dirac operator, and the
second factor is due to the presence of $|\nu|$ exact zero modes.  The
partition function for fixed topological charge $\nu$ is given by the
average of the fermion determinant over gauge-field configurations
with fixed $\nu$. Defining
\begin{align}
  \hat Z_{|\nu|}(|M|)
  =\Bigl\langle\prod_{\lambda_n>0}\det(\lambda_n^2+MM^\dagger)\Bigr\rangle_\nu\,,
\end{align}
which only depends on $|M|=\diag(|m_1|,\ldots,|m_{N_f}|)$ and $|\nu|$, the
partition function reads
\begin{align}\label{eq:Znu}
  Z_\nu(M)=\hat Z_{|\nu|}(|M|)\times
  \begin{cases}
    \det^\nu M, & \nu\ge0\,,\\
    \det^{-\nu} M^\dagger, & \nu<0\,.
  \end{cases}
\end{align}
The partition function at fixed $\theta$-angle is then given by
\begin{align}
  \label{eq:sumnu}
  Z(M,\theta) &= \sum_{\nu = -\infty}^\infty e^{i\nu\theta}Z_\nu(M) \notag\\
  & = \sum_{\nu = -\infty}^\infty
  e^{i\nu\big(\theta+\sum\limits_{k=1}^{N_f}\phi_k\big)}
  \det |M|^{|\nu|} \hat Z_{|\nu|}(|M|)\,,
\end{align}
where $m_k=|m_k|e^{i\phi_k}$ defines the phase $\phi_k$ of $m_k$. It is
clear that $Z(M,\theta)$ is a periodic function of $\theta$, and if
$\sum_k\phi_k$ is a multiple of $\pi$ it is also even in $\theta$.  It only depends
on the sum of the phases of the quark masses, and this sum can be
absorbed in a redefinition of the $\theta$-angle. The same statement is
true for the spectral density, whose mass dependence comes only from
the fermion determinant in the statistical measure.

Therefore, from a mathematical point of view, it suffices to derive
results for real and non-negative quark masses.  However, we will
sometimes consider the ``physical" situation where one of the quark
masses is taken to be negative. This case can be obtained by shifting
$\theta\to\theta+\pi$ in the mathematical result. Nevertheless, our results assume
$m_k\ge0$ unless stated otherwise.

The chiral condensate of flavor $k$ is defined as 
\begin{align}
  \label{eq:Sigmam}
  \Sigma(m_k)&=-\langle \bar \psi_k \psi_k \rangle
  =-\langle \bar \psi_{kR} \psi_{kL} \rangle
  -\langle \bar \psi_{kL} \psi_{kR} \rangle \notag\\
  &=\frac1{N_dV}\left(\frac d{dm_k}+\frac d{dm_k^*}\right)\log Z(M)\,,
\end{align}
where we have suppressed the dependence of $\Sigma(m_k)$ on $M$ and where
$N_d$ (with $d$ for ``degenerate'') is the number of quarks whose mass
equals $m_k$. Note that Eq.~\eqref{eq:Sigmam} is valid both at fixed
$\nu$ and at fixed $\theta$. Let us make two remarks here.  First, for a real
and negative mass, the derivatives in Eq.~\eqref{eq:Sigmam} simply
lead to an extra sign (compared with the result for a positive
mass). Second, for a genuinely complex mass $m_k=|m_k|e^{i\phi_k}$, one
can rewrite the derivatives in terms of $|m_k|$ and $\phi_k$. For the
partition function~\eqref{eq:sumnu} at fixed $\theta$ the derivative
w.r.t.\ $\phi_k$ can be rewritten as a derivative w.r.t.\ $\theta$.

Let us begin with the case of one flavor of mass $m$. Since the free
energy is extensive in $V$, the QCD partition function, obtained by
expanding the action to lowest order in $m$, is given by
\cite{Leutwyler:1992yt}
\begin{align}\label{one}
  Z(m,\theta) = Z(m=0,\theta)\exp[mV \Sigma \cos\theta +O(m^2 V)]\,,
\end{align}
where $\Sigma$ is the absolute value of the chiral condensate in the limit
$m=0$ and $\theta=0$. Since the mass-independent factor
$Z(m=0,\theta)$ does not contribute to the chiral condensate we will ignore
it below.  Equation~\eqref{one} is valid for both positive and
negative quark mass.  This has the consequence that the chiral
condensate from Eq.~\eqref{eq:Sigmam},
\begin{align}
\label{eq:Sigmaonef}
  \Sigma(m) = \Sigma \cos \theta\,,
\end{align}
does not change sign when $m$ becomes negative.  Assuming nonvanishing
$\Sigma$, at first sight this appears to be in contradiction with the
Banks-Casher formula \cite{Banks:1979yr}, which expresses the chiral
condensate in terms of the eigenvalues $\lambda_n$ of the Dirac operator
starting from the relation
\begin{align}\label{eq:Sigmalambda}
  \Sigma(m) = \biggl
  \langle \frac 1V \sum_n \frac 1{i\lambda_n+m}\biggr \rangle\,,
\end{align}
where the average is over gauge-field configurations.
Equation~\eqref{eq:Sigmalambda} is valid both at fixed $\nu$ and at
fixed $\theta$. The eigenvalues are either zero, giving a term proportional
to $1/m$, or they occur in pairs $\pm\lambda_n$, which yield terms of the
form $2m/(\lambda_n^2+m^2)$. Thus the function $\Sigma(m)$ appears to be odd in
$m$. The resolution of this puzzle is that the statistical weight in
Eq.~\eqref{eq:Sigmalambda} contains the fermion determinant, which
leads to an additional mass dependence.  This resolution was fully
worked out in Ref.~\cite{Verbaarschot:2014upa,Verbaarschot:2014qka} by
an explicit computation of the spectral density and the chiral
condensate in the $\epsilon$-domain of QCD. Employing identities for sums of
products of Bessel functions, the expressions could be summed to give
the spectral density at $\theta=0$.  For negative mass the resulting
expression is increasing exponentially with $V$ and oscillating with a
period that scales as $1/V$.  As we know from QCD at nonzero chemical
potential \cite{Osborn:2005ss}, exactly such behavior of the spectral
density can eliminate a discontinuity of the chiral
condensate. However, as has already been observed in
\cite{Kanazawa:2011tt}, the contributions of both the zero modes and
the nonzero modes diverge exponentially with the volume. It turns out
\cite{Verbaarschot:2014upa,Verbaarschot:2014qka} that these divergent
contributions cancel identically, resulting in a chiral condensate
that remains constant in the $\epsilon$-domain, i.e., for
$m\Lambda_\text{QCD}\sqrt V \ll 1$.
  
Let us turn to two-flavor QCD. In this case the full flavor symmetry
is $\U(2)\times\U(2)$, with the axial $\U(1)$ group broken by the anomaly
and the $\SU(2)\times\SU(2)$ subgroup broken spontaneously by the chiral
condensate.  In the $\epsilon$-domain of QCD, the partition function of the
resulting Nambu-Goldstone modes, which interact according to a chiral
Lagrangian, simplifies to \cite{Leutwyler:1992yt}
\begin{align}\label{eq:Z}
  Z(M,\theta) = \int\limits_{\SU(N_f)}dU\exp\bigl[V\Sigma
  \re(e^{i\theta/N_f}\Tr MU)\bigr]\,,
\end{align}
which actually holds for any $N_f$.  The measure $dU$ is the
normalized Haar measure.  In the thermodynamic limit, the $U$ field
aligns itself with the chiral condensate. For $N_f=2$, the simplest
case is when the two masses are equal to a common mass $m$. For
$m\cos(\theta/2)>0$ the saddle-point solution is $U = \1$, but for
$m\cos(\theta/2)<0$ it is given by $U = -\1$. To leading order in the
thermodynamic limit this results in
\begin{align}\label{eq:ZNf2}
  \log Z(m,m,\theta) \overset{|m|V\Sigma\gg 1}{\approx} 2 V \Sigma |m\cos(\theta/2)|\,.
\end{align}
Because of the absolute value, the chiral condensate as defined in
Eq.~\eqref{eq:Sigmam} acquires a discontinuity at $m = 0$,
\begin{align}
  \Sigma(m)\overset{|m|V\Sigma\gg 1}{\approx}\sign(m)\Sigma|\cos(\theta/2)|\,.
\end{align}
\begin{figure}[t!]
  \centerline{\includegraphics[width=.95\columnwidth]{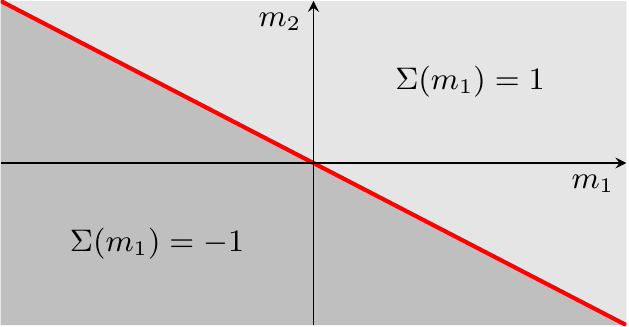}}
  \caption[]{Phase diagram of the two-flavor theory at $\theta=0$ in
    the plane of the two quark masses.}
  \label{fig2}
\end{figure}

Another simple case is that of unequal quark masses and $\theta=0$, where
we have in leading order \cite{Leutwyler:1992yt}
\begin{align}
  \log Z(m_1,m_2,\theta=0)\overset{|m_k|V\Sigma\gg 1}{\approx}V\Sigma|m_1+m_2|\,.
\end{align}
The phase diagram of this case is shown in Fig.~\ref{fig2}. The two
phases are separated by the line $m_1+m_2 = 0$ on which the pion mass
becomes zero, known as the Dashen phenomenon \cite{Dashen:1970et}.
For three flavors, this line changes to a finite region where pions
condense and CP symmetry is spontaneously broken
\cite{Creutz:2003xu}. It can also become a finite region when pion
condensation occurs in the case of Wilson fermions
\cite{Horkel:2014nha}.

\begin{figure}[b]
  \includegraphics[width=\columnwidth]{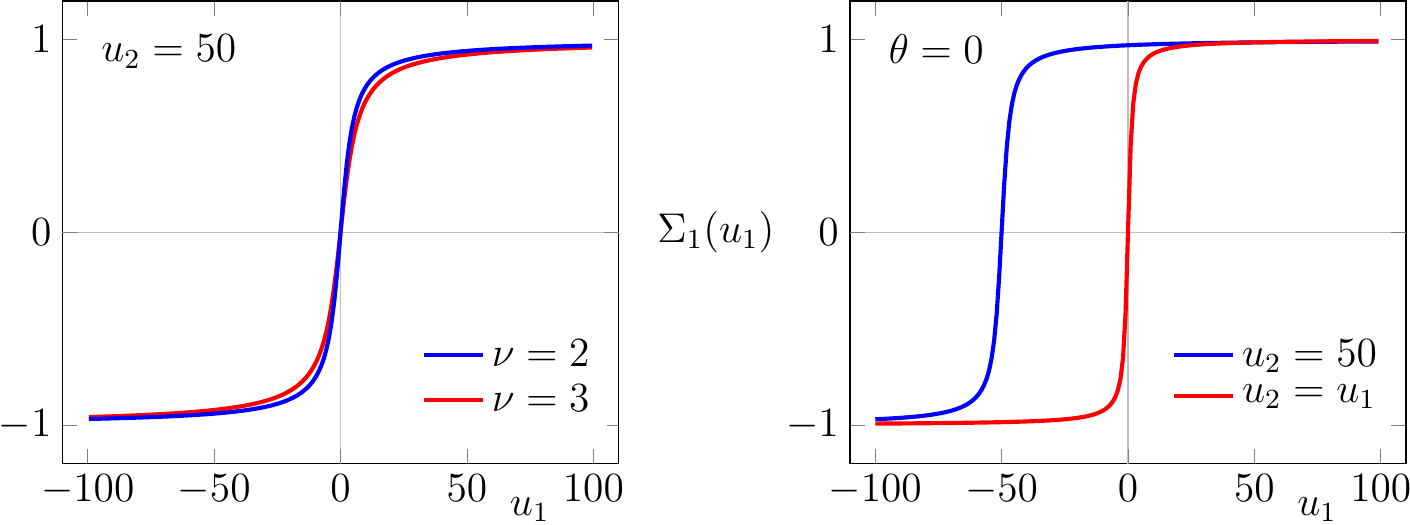}
  \caption[]{Mass dependence of the chiral condensate for two-flavor
    QCD at fixed topological charge $\nu$ (left) and at fixed
    $\theta = 0$ (right).  In the thermodynamic limit all curves become
    discontinuous.  Note that we use the dimensionless masses
    $\m_k=m_kV\Sigma$. The notation $\Sigma_1$ indicates that we differentiate
    w.r.t.\ the first quark mass $\m_1$.}
  \label{fig:2}
\end{figure}

The phase diagram at fixed $\theta$ has to be contrasted to the one at
fixed topological charge $\nu$, where we have in leading order
\cite{Leutwyler:1992yt}
\begin{equation}
  \log Z_\nu(m_1,m_2)\overset{|m_k|V\Sigma\gg 1}{\approx}V\Sigma(|m_1|+|m_2|)\,.
\end{equation}
The sign of the chiral condensate at fixed $\nu$ always changes sign
when one of the masses changes its sign, unlike at fixed $\theta=0$, where
the sign change occurs at the Dashen line $m_1=-m_2$. One aim of the
present work is to understand this difference. In Fig.~\ref{fig:2} we
highlight the different behavior of the two-flavor chiral condensate
at fixed topological charge $\nu$ (left) and fixed $\theta$-angle (right).

Finally, let us mention one particular limit of the two-flavor
case. When one of the quark masses becomes large, the two-flavor
theory reduces to the one-flavor theory. In the chiral Lagrangian we
then have
\begin{align}
  U \to 
  \begin{pmatrix}
    e^{-i\phi} &0 \\ 0 &  e^{i\phi}    
  \end{pmatrix}
\end{align}
with $\phi$ fixed at $\phi = \theta/2$, resulting in the one-flavor partition
function~\eqref{one}.

\section{ Sign Problem}
\label{sec:sign}

Generically, QCD at nonzero $\theta$-angle has a sign problem, which
originates from the weight factor $\exp[i\nu(\theta+\sum_k\phi_k)]$ in
Eq.~\eqref{eq:sumnu}. The sign problem is absent only for
$\cos(\theta+ \sum_k\phi_k)=1$, and it is most severe for
$\cos(\theta+ \sum_k\phi_k)=-1$, in which case the weight factor is
$(-1)^\nu$. For real (but possibly negative) quark masses, these two
conditions translate into $\cos\theta=\pm\sign\det M$.

While these two extreme cases are already apparent from
Eq.~\eqref{eq:sumnu}, the severity of the sign problem in the general
case can be measured by the ratio of the partition function with a
phase and the phase-quenched partition function. Explicit analytical
results for this ratio can be obtained in the $\epsilon$-domain.

\begin{figure}[t!]
  \centerline{\includegraphics[width=.9\columnwidth]{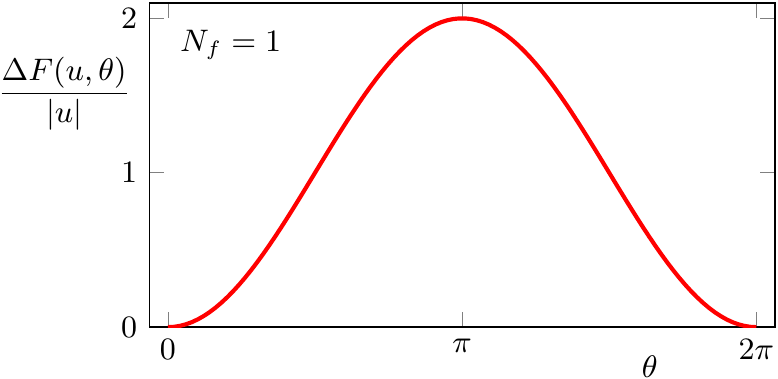}}
  \caption{Difference of the free energies of the phase-quenched and
    the full theory for one-flavor QCD, see
    Eq.~\eqref{free-energy-Nf1}, divided by the absolute value of the
    rescaled quark mass. Here $\m>0$.}
  \label{fig:freeNf1}
\end{figure}

For one flavor the ratio is given by
\begin{align}\label{free-energy-Nf1}
  \exp[- \Delta F(m,\theta)]=\frac{Z(m,\theta)}{Z(|m|,0)} 
  =e^{V\Sigma (m\cos\theta-|m|)}\,.
\end{align}
Therefore exponential cancellations take place at $\theta\ne0$.  The function
$\Delta F(m,\theta)=V\Sigma|m|(1-\sign(m)\cos\theta)$ is shown in Fig.~\ref{fig:freeNf1}
for $m>0$.  In agreement with the general argument above, it assumes
its maximum at $\cos\theta=-\sign(m)$, where the sign problem is most
severe, while the sign problem is absent for $\cos\theta=\sign(m)$. The
free energy is a smooth function of the quark mass and the
$\theta$-angle, which reflects the fact that the one-flavor theory has no
phase transition.
 
\begin{figure}
  \centerline{\includegraphics[width=\columnwidth]{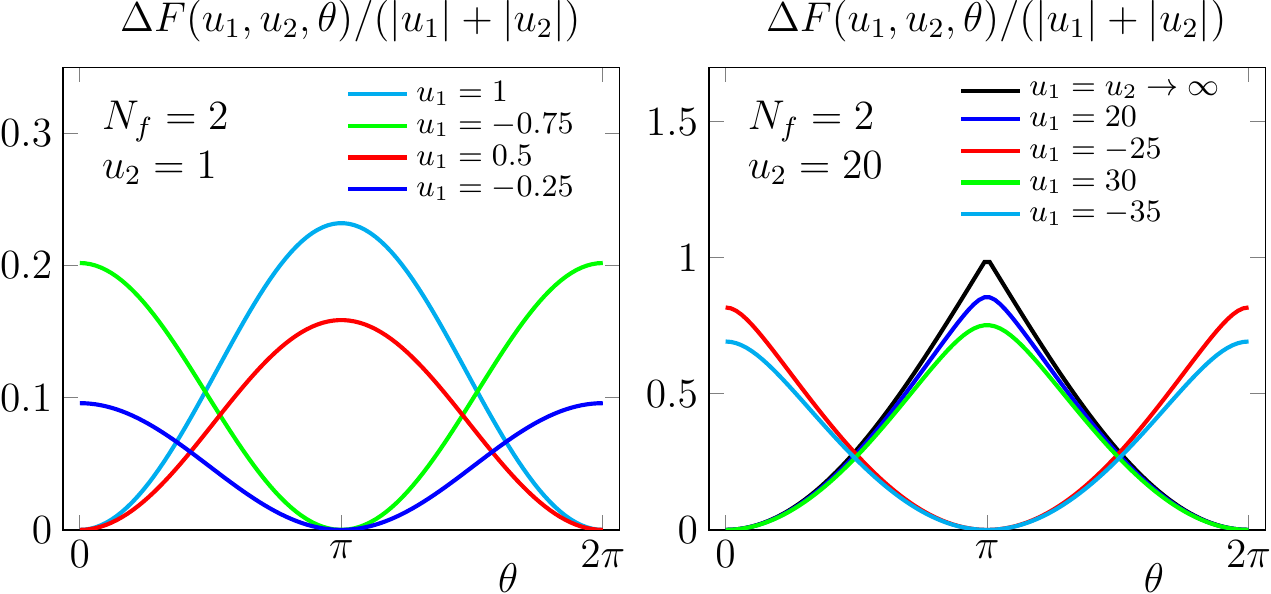}}
  \caption{Free-energy difference as a function of $\theta$ for two flavors
    and several quark masses. The functions are normalized by the sum
    of the quark masses so that the large-mass limit remains
    finite. The black curve in the plot on the right is the
    thermodynamic limit for equal quark masses, see
    Eq.~\eqref{free-Nf2-limit}.  }
  \label{fig:freeNf2}
\end{figure}

For two flavors the ratio is given by
\begin{align}\label{free-energy-Nf2}
  \exp[- \Delta F(m_1,m_2,\theta)]=\frac{Z(m_1,m_2,\theta)}{Z(|m_1|,|m_2|,0)}\,.
\end{align}
The $\epsilon$-domain result for the two-flavor partition function is given
by the integral over $\SU(2)$ in Eq.~\eqref{eq:Z}. In terms of the
microscopic variables $\m_k=m_kV\Sigma$ it becomes \cite{Leutwyler:1992yt}
\begin{align}\label{eq:Z2}
  Z(\m_1,\m_2,\theta) =\frac {2I_1\bigl(\sqrt{\m_1^2+\m_2^2+2\m_1 \m_2 \cos\theta}\bigr)}
  {\sqrt{\m_1^2+\m_2^2+2\m_1 \m_2 \cos\theta}}
\end{align}
with $I_k$ the modified Bessel function of the first kind.  Again the
difference $\Delta F$ assumes its maximum at
$\cos\theta=-\sign(m_1 m_2)$, see the plots in
Fig.~\ref{fig:freeNf2}. Therefore the sign problem is most severe
either at $\theta=0$ when both masses have opposite signs, or at
$\theta=\pi$ when both masses have the same sign. This can also be seen in
the exponentially increasing oscillations of the level density when
increasing the $\theta$-angle, see Fig.~\ref{fig:rhoNf2} below. The sign
problem is absent only for $\cos\theta=\sign(m_1 m_2)$.

We illustrate the free-energy difference in
Fig.~\ref{fig:freeNf2}. The partition function \eqref{eq:Z2} is
monotonically decreasing (increasing) in $\theta\in[0,\pi]$ for quark masses of
equal (opposite) signs. This carries over to an increase (decrease) of
the free energy. The free-energy difference is also strictly
increasing (decreasing) with respect to the moduli of the masses for
equal (opposite) signs. For equal masses it simplifies to
\begin{equation}\label{free-energy-Nf2.b}
  \Delta F(\m,\m,\theta)=\log\left[\frac{I_1(2\m)\cos(\theta/2)}{I_1(2\m\cos(\theta/2))}\right],
\end{equation}
where the monotonicity can be checked easily.

The free-energy difference has the thermodynamic limit
\begin{align}\label{free-energy-Nf2-limit}
  &\lim_{|\m_1|,|\m_2|\gg1}\Delta F(\m_1,\m_2,\theta)\notag\\
  &\qquad=|\m_1|+|\m_2|-\sqrt{\m_1^2+\m_2^2+2\m_1\m_2\cos\theta}\,,
\end{align}
where only the Bessel functions had to be approximated. (There are
subleading logarithmic corrections to this result.) For equal masses
the thermodynamic limit becomes
\begin{equation}\label{free-Nf2-limit}
  \lim_{|\m|\gg1}\Delta F(\m,\m,\theta)=2|\m|\left(1-\left|\cos(\theta/2)\right|\right),
\end{equation}
which has a kink at the Dashen point $\theta=\pi$.

\section{Decomposition of the Spectral Density and the Chiral
  Condensate}
\label{sec:decom}

In the limit of zero quark masses, the chiral condensate at fixed
topological charge has a discontinuity on the support of the Dirac
spectrum, which is dense in the thermodynamic limit.  We have to
understand how this discontinuity can be moved away from the support
of the spectrum at nonzero $\theta$-angle. When the spectral density is
positive this is certainly not possible. However, the averaging
procedure to obtain the spectral density includes the fermion
determinant.  This determinant is not positive definite at nonzero
$\theta$-angle, and thus the spectral density is generically not positive
definite.  Moreover, the spectral density is normalized with respect
to the partition function at $\theta \ne 0$, which is exponentially smaller
than the phase-quenched partition function, and therefore may result
in a spectral density that increases exponentially with the
volume. Indeed, we know from QCD at nonzero chemical potential
\cite{Osborn:2005ss} and from QCD-like theories with indefinite
measure \cite{Kanazawa:2011tt} that the discontinuity of the chiral
condensate can be shifted when the spectral density oscillates with an
amplitude that is exponentially large in $V$ and with a period that
scales as $1/V$. As a first step towards understanding this behavior,
we decompose the spectral density and the chiral condensate into
various contributions.

We denote the spectral density of the Dirac operator at fixed
topological charge $\nu$ by $\rho_\nu(\lambda,M)$. The spectral density at fixed
$\theta$ is defined by
\begin{align}
  \label{eq:rhotheta}
  \rho(\lambda,M,\theta) = \sum_\nu  P_\nu(M,\theta) \rho_\nu(\lambda,M)\,,
\end{align}
where $P_\nu$ is the statistical weight to find a gauge-field
configuration with topological charge $\nu$,
\begin{align}
  \label{eq:Ptheta}
  P_\nu(M,\theta) = \frac{Z_\nu(M) e^{i\nu\theta} }{\sum_\nu Z_\nu(M) e^{i\nu\theta}}
  = \frac{Z_\nu(M) e^{i\nu\theta} }{Z(M,\theta)}\,.
\end{align}
To obtain a more detailed picture we split the spectral density into a
zero-mode part and a nonzero-mode part,
\begin{align}\label{eq:split}
  \rho(\lambda,M) = \rho^\zm(\lambda,M) + \rho^\nzm(\lambda,M)\,.
\end{align}
This splitting is valid both at fixed $\nu$ and at fixed $\theta$.
Equation~\eqref{eq:rhotheta} holds separately for the zero-mode and
nonzero-mode parts.

The zero-mode part of the density at fixed $\nu$ is
\begin{align}
  \label{eq:rhoz}
  \rho^\zm_\nu(\lambda)=|\nu|\delta(\lambda)\,.
\end{align}
We will see in the next section that $\rho^\zm$ gives a contribution to
the chiral condensate that diverges exponentially if a sign problem is
present. This contribution must be canceled by a similar contribution
of the nonzero-mode part $\rho^\nzm$ to obtain a finite condensate. The
question is what part of $\rho^\nzm$ is responsible for this
cancellation. Obviously there is no unique answer to this question,
but we know that this cancellation also has to take place in the
quenched approximation.  Therefore we decompose $\rho^\nzm$ at fixed
$\nu$ into a quenched part (obtained by setting $N_f=0$) and a dynamical
part (the remainder),
\begin{align}
  \label{eq:splitq}
  \rho_\nu^\nzm(\lambda,M)=\rho_\nu^\q(\lambda)+\rho_\nu^\d(\lambda,M)\,.
\end{align}
The $\epsilon$-domain result for the quenched part in terms of the
microscopic variable $\x=\lambda V\Sigma$ reads
\cite{Verbaarschot:1993pm}
\begin{align}
  \label{eq:rhoq}
  \rho_\nu^\q(\x)=\frac{|x|}2\big[J_\nu^2(x)-J_{\nu+1}(x)J_{\nu-1}(x)\big]
\end{align}
with $J_k$ the Bessel function of the first kind.  The dynamical part
depends, in addition to $\x$, also on the quark masses, which on the
microscopic scale we collect in $\M=(\m_1,\ldots,\m_{N_f})$. Explicit
expressions can be found in \cite{Damgaard:1997ye,Wilke:1997gf}.

An alternative to Eq.~\eqref{eq:splitq} would be to split $\rho^\nzm$ at
fixed $\theta$ into a phase-quenched part \cite{Verbaarschot:2014qka},
obtained by letting $M\to|M|$ and $\theta\to0$, and an oscillating
remainder. The cancellation is then achieved by the phase-quenched
part, but this part is more complicated than the quenched part, and
therefore we do not consider it in this paper.

The chiral condensate is obtained from the spectral density by the
relation
\begin{align}\label{eq:sigma}
  \Sigma(m,M) = \frac1V\int_{-\infty}^\infty d\lambda\, 
  \frac{\rho(\lambda,M)}{i\lambda+m}\,,
\end{align}
where $m$ is a valence quark mass. It is convenient to distinguish $m$
from the sea quark masses in $M$, but at the end of the calculation
$m$ will usually be set equal to one of the sea quark masses. Again,
Eq.~\eqref{eq:sigma} holds both at fixed $\nu$ and at fixed $\theta$. On the
macroscopic scale Eq.~\eqref{eq:sigma} requires regularization, but in
the $\epsilon$-domain it is valid as it stands.

We now split the condensate into zero-mode, quenched and dynamical
part, obtained by replacing $\rho$ in Eq.~\eqref{eq:sigma} by
$\rho^\zm$, $\rho^\q$ and $\rho^\d$, respectively. This leads to
\begin{align}
  \Sigma(m,M) &= \Sigma^\zm(m,M) + \Sigma^\q(m,M) + \Sigma^\d(m,M)
                        \label{eq:splitSigma}
\end{align}
at either fixed $\nu$ or fixed $\theta$.  As in Eq.~\eqref{eq:rhotheta} we
have
\begin{align}
  \Sigma(m,M,\theta) = \sum_\nu P_\nu(M,\theta) \Sigma_\nu(m,M)\,,
\end{align}
and this holds separately for all three contributions.

All equations in this section can be translated to the microscopic
scale by the replacements $\lambda=x/V\Sigma$ and $M=\M/V\Sigma$. In
Eq.~\eqref{eq:sigma}, the prefactor $1/V$ is then replaced by
$\Sigma$. For the valence mass on the microscopic scale we will use the
notation $\hu=mV\Sigma$.

\section{Cancellation of zero-mode and quenched contribution}
\label{section-4}

In this section we show, for any number of flavors, that the
exponentially increasing contribution of the zero modes to the chiral
condensate is canceled by an exponentially increasing contribution
from the quenched part of the spectrum.  The expressions for the
contribution of the zero modes are valid without any assumptions, but
the other calculations are performed in the $\epsilon$-domain of QCD.

Using Eqs.~\eqref{eq:rhotheta} and \eqref{eq:rhoz}, the zero-mode part
of the microscopic spectral density at fixed $\theta$ is given by
\begin{align}
  \rho^\zm(\x,\M,\theta)=\frac{\delta(\x)}{Z(\M,\theta)} \sum_\nu e^{i\nu\theta} |\nu| Z_{\nu}(\M)\,.
\label{sub1}
\end{align}
Using the Fourier transform of Eq.~\eqref{eq:sumnu},
\begin{align}
  \label{eq:inverse}
  Z_\nu(\M)=\int_{-\pi}^\pi \frac{d\theta}{2\pi}\,e^{-i\nu\theta}Z(\M,\theta)\,,
\end{align}
the sum on the RHS can be rewritten in the form
\begin{align}
  \sum_\nu e^{i\nu\theta} |\nu| Z_{\nu}(\M)
  =\dashint_{-\pi}^\pi \frac {d\phi}{2\pi}\, A(\phi) \big[Z(\M,\theta-\phi)-Z(\M,\theta)\big]\,,
  \label{zm1}
\end{align}  
where $A(\phi)$ is the Fourier transform
\begin{align}
  \label{eq:Aphi}
  A(\phi) =\sum_\nu e^{i\nu\phi} |\nu|
\end{align}
and the symbol $\dashint$ stands for a principal-value integral.
Several comments are in order. (i) The definition \eqref{eq:Aphi} is
to be understood as a distribution acting on test functions that are
twice differentiable and vanish at $\phi =0$, where the sum over $\nu$ is
divergent. We always assume $\phi\in[-\pi,\pi]$ because of the
integral~\eqref{zm1}. (ii) Since the integral over $A(\phi)$ vanishes, we
could subtract $Z(\M, \theta)$ in \eqref{zm1} to end up with a test
function that indeed vanishes at $\phi=0$. (iii) In the following we
always deal with test functions that are twice differentiable and
vanish at $\phi=0$. This justifies the introduction of a regulator
$\epsilon>0$ which results in a sum that is point wise convergent except for
$\phi = 0$,
\begin{align}
  A(\phi) &=\lim_{\epsilon\to 0}\sum_{\nu=-\infty}^\infty |\nu|e^{i\nu\phi-|\nu|\epsilon}\notag\\
  &=\lim_{\epsilon\to 0}\sum_{\nu=1}^\infty \nu (e^{-(\epsilon+i\phi)\nu}+e^{-(\epsilon-i\phi)\nu})\notag\\
  &=\lim_{\epsilon\to 0}\frac\partial{\partial\epsilon}\frac{e^{-\epsilon}-\cos\phi}{\cosh\epsilon-\cos\phi}\notag\\
  &=-\frac 1{2\sin^2\frac\phi2} \quad(\text{valid for }\phi\ne0)\,.
\label{adelt}
\end{align}
Note that due to the point wise convergence, the integral of $A(\phi)$ in
the form \eqref{adelt} no longer vanishes so that the subtractions
have to be made before the limit $\epsilon\to 0$ is taken.  For a function
$f$ that is twice differentiable at zero we therefore have
\begin{align}
  \label{eq:Aphi.2}
    \dashint_{-\pi}^\pi d\phi\,A(\phi)\big[f(\phi)-f(0)\big]
    =-\dashint_{-\pi}^\pi d\phi\,\frac{f(\phi)-f(0)}{2\sin^2\frac\phi2}\,.
\end{align}
Finally, we note that
$-1/2\sin^2\tfrac\phi2=\partial_\varphi\cot\frac\varphi2$ is a total derivative which can be
integrated by parts on test functions that vanish at $\phi = 0$.

The zero-mode part of the spectral density thus becomes
\begin{align}
  \label{eq:rhozm}
  \rho^\zm(\x,\M,\theta) =\delta(\x) \dashint_{-\pi}^\pi
  \frac {d\phi}{2\pi}\frac {1}{2\sin^2\frac\phi2}
  \left[ 1-\frac{Z(\M,\theta\!-\!\phi)}{Z(\M,\theta)}\right ].
\end{align}
The zero-mode contribution to the chiral condensate follows trivially
from Eq.~\eqref{eq:sigma},
\begin{align}
  \label{eq:sigmazm}
  \frac{\Sigma^\zm(\hu,\M,\theta)}{\Sigma} = \frac 1{\hu} \dashint_{-\pi}^\pi
  \frac {d\phi}{2\pi}\frac {1}{2\sin^2\frac\phi2} 
\left [ 1- \frac{Z(\M,\theta-\phi)}{Z(\M,\theta)}\right].
\end{align}
In the derivation of this formula we did not make any assumptions on
the scaling behavior of the quark masses, and we stress that this
result is valid in general.

From the discussion of the sign problem in Sec.~\ref{sec:sign} we know
that the ratio $Z(\M,\theta)/Z(|\M|,0)$ takes its maximum at
$\cos\theta=\sign\det M$. Hence, for $\det M >0$, the
integral~\eqref{eq:sigmazm} is dominated by the region around
$\phi=\theta$ , and for $\det M <0$ it is dominated by the region around
$\phi=\theta-\pi$.  For $\cos\theta\neq\sign\det M$, the partition function
$Z(\M, \theta)$ is exponentially smaller (in $\M$) than
$Z(|\M|, \theta =0)$, resulting in a contribution to the chiral condensate
that increases exponentially with the volume.

The quenched part of the spectral density at fixed $\theta$ is
\begin{align}
  \rho^{q}(\x,\M,\theta) &= \frac {1}{Z(\M,\theta)} \sum_\nu e^{i\nu\theta}
                        \rho^{q}_\nu(x)  Z_{\nu}(\M) \notag\\
  &= \int_{-\pi}^\pi \frac {d\phi}{2\pi}\, \tilde \rho^\q (x,\phi)
    \frac{Z(\M,\theta-\phi)}{Z(\M,\theta)}
  \label{eq:rhoqtheta}
\end{align}
with
\begin{align}
  \label{eq:rhoqtilde}
  \tilde \rho^\q(\x,\phi) = \sum_\nu e^{i\nu\phi}\rho^\q_\nu(\x)
  = \frac{J_1(2|\x|\sin\frac\phi2)}{2\sin\frac\phi2}\,.
\end{align}
The last equality in Eq.~\eqref{eq:rhoqtilde} was obtained from
Eq.~\eqref{eq:rhoq} using Eq.~\eqref{a1.5}. To avoid confusion, we
note that $\tilde\rho^\q(\x,\theta)$ is the spectral density of the true
quenched theory at fixed $\theta$. It is convoluted with
$Z(\M,\theta-\phi)/Z(\M,\phi)$ as shown in Eq.~\eqref{eq:rhoqtheta} to obtain
what we have defined as the quenched part of the spectral density. The
convolution introduces the quark masses into $\rho^\q(\x,\M,\theta)$, while
$ \tilde \rho^\q(\x,\phi)$ is independent of the quark masses.

Equation~\eqref{eq:rhoqtheta} results in the ``quenched" contribution
to the chiral condensate
\begin{align}
  \label{sigqgen}
  &\frac{\Sigma^\q(\hu,\M,\theta)}\Sigma \\
  &= \int_0^\infty  \frac{2d\x\,\hu}{\x^2+\hu^2}
    \int_{-\pi}^\pi \frac {d\phi}{2\pi}
    \frac{J_1(2\x\sin\frac\phi2)}{2\sin\frac\phi2} 
    \frac{Z(\M.\theta-\phi)}{Z(\M,\theta)}\notag\\
  &=\int_{-\pi}^\pi \frac {d\phi}{2\pi}
    \left[\frac 1{2\hu\sin^2\frac\phi2} - \frac{ \Khuphi1}
    {\sign(\hu)|\sin\frac\phi2|}\right ]
    \frac{Z(\M,\theta-\phi)}{Z(\M,\theta)}\,.\notag
\end{align}
In the last line we have used Eq.~\eqref{eq:J11} to obtain the
modified Bessel function of the second kind $K_1$. Note that the poles
at $\phi=0$ cancel so that the integral can be evaluated as an ordinary
integral.  Also for the quenched contribution, the integral is
dominated by the region around $ \phi = \theta$ or $\phi=\theta-\pi$ (depending on the
sign of the product of the quark masses) where the effective
$\theta$-angle vanishes.  For $\theta+\arg\det M \ne 0$, this again leads to
contributions that increase exponentially. For $N_f=0$ the ratio of
partition functions in Eq.~\eqref{sigqgen} is equal to unity. The
integral over $\phi$ of the expression in square brackets is equal to the
``quenched'' part of the condensate at topological charge zero, i.e.,
$\Sigma_{\nu=0}^\q(\hu)$, which follows by writing the RHS of
Eq.~\eqref{eq:rhoqtilde} as a Fourier sum.

We now show that the two exponentially increasing contributions to the
chiral condensate cancel.  The sum of zero-mode and quenched
contribution is given by
\begin{align}
  \label{sigsum0}
  &\frac{\Sigma^\zm(\hu,\M,\theta)}\Sigma+\frac{\Sigma^\q(\hu,\M,\theta)}\Sigma \\
  &=\dashint_{-\pi}^\pi\frac{d\phi}{2\pi}
    \left [\frac 1{2\hu \sin^2\frac\phi 2}
    - \frac{\Khuphi1)}{\sign(\hu)|\sin\frac\phi2|}
    \frac{Z(\M,\theta\!-\!\phi)}{Z(\M,\theta)}\right]\notag\\
  &=\frac{\Sigma_{\nu=0}^\q}\Sigma -
    \dashint_{-\pi}^\pi\frac{d\phi}{2\pi}
    \frac{\Khuphi1)}{\sign(\hu)|\sin\frac\phi2|}
    \left[\frac{Z(\M,\theta-\phi)}{Z(\M,\theta)}-1\right].\notag
\end{align}
This integral converges as a principal-value integral.

The partition function is given in Eq.~\eqref{eq:Z}. We write
$U\in\SU(N_f)$ in the form
$U=V\diag(e^{i\phi_1},\ldots,e^{i\phi_{N_f}})V^\dagger$ with
$V\in\U(N_f)/\U(1)^{N_f}$ and $\sum_k\phi_k=0$ and perform a change of
variables from $U$ to $V$ and $\vec\phi=(\phi_1,\ldots,\phi_{N_f})$. The Jacobian of
this transformation is independent of $V$. If we assume degenerate
quark masses, the integrand is also independent of $V$ so that the
integration over $V$ simply gives an irrelevant constant. For
non-degenerate quark masses the integral over $V$ is of the
Harish-Chandra--Itzykson-Zuber type \cite{Itzykson:1979fi} and leads to
\begin{align}
  Z(\M,\theta) &\sim \int \prod_k \frac{d\phi_k}{2\pi}\,
   |\Delta(e^{i\vec\phi})|^2 
     \delta\Big(\sum_k\phi_k\Big)\notag\\
     &\quad\times\frac{ \det \left[\exp[ \m_k\cos(\phi_\ell +\theta/N_f)]\right]}
   {\Delta(\M)\Delta(\cos(\vec\phi+\theta/N_f))}\,,
   \label{zu0}
\end{align}
where $\Delta(\vec x)=\prod_{a<b}(x_b-x_a)$ denotes the Vandermonde
determinant.  Unless indicated otherwise, a product or a sum over $k$
is understood to run from 1 to $N_f$.  The symbol $\sim$ indicates that
we have suppressed the normalization constant.
 
The integrand in~\eqref{zu0} is symmetric under permutations of the
$\phi_k$. We denote the non-exponential terms by
$f(\M,\vec\phi+\theta/N_f)= |\Delta(e^{i\vec\phi})|^2 /(\Delta(\M)\Delta(\cos(\vec\phi+\theta/N_f)))$
and expand the determinant of $\exp[ \m_k\cos(\phi_\ell+\theta/N_f)]$ to
obtain
\begin{align}
  Z(\M,\theta) &\sim \int \prod_k \frac{d\phi_k}{2\pi}\,
  f\left(\M,\vec\phi+\theta/N_f\right)\notag\\
  &\quad\times e^{\sum\limits_k \m_k\cos(\phi_k +\theta/N_f)}\delta\Big(\sum_k\phi_k\Big)\,.
  \label{zu}
\end{align}
After shifting each $\phi_k$ by $-\theta/N_f$ and performing the integration
over $\phi_1$ using the $\delta$-function, Eq.~\eqref{zu} becomes
\begin{align}\label{zmth}
  Z(\M, \theta) &\sim \int \prod_{k\ge2}\frac {d\phi_k}{2\pi}
  f\bigl(\M,\vec{\bar\phi}\bigr)\\
  &\quad\times  \exp\Bigl[\m_1\cos\Bigl(\theta-\sum\limits_{k\ge2}\phi_k\Bigr)+\sum\limits_{k\ge2}\m_k\cos\phi_k\Bigr]\,.\notag
\end{align}
where $\vec{\bar\phi}=(\theta-\sum_{k\geq2}\varphi_k,\phi_2,\ldots,\phi_k)$. We will use this
representation for the partition function in the denominator of
Eq.~\eqref{sigsum0}.
 
Next, we consider the integral in the last line of
Eq.~\eqref{sigsum0}. For large $\hu$, the Bessel function behaves as
\begin{equation}
  \label{eq:Kasympt}
  \Khuphi\nu=
  \frac{\sqrt \pi}{\sqrt{4|\hu\sin\frac\phi2|}} e^{-2|\hu\sin\frac\phi2|}
  \left[1+O\left(1/\hu\right)\right].
\end{equation}
Therefore the $-1$ term in the last line of Eq.~\eqref{sigsum0} cannot
result in exponentially large contributions, but this term regularizes
the integral at $\phi =0$.


We now combine Eq.~\eqref{zu} with $\theta$ replaced by
$\theta-\varphi$ and Eq.~\eqref{eq:Kasympt} and consider the contribution to the
chiral condensate that gives the exponentially large terms.  After
shifting each $\varphi_k$ by $(\varphi-\theta)/N_f$, we obtain
\begin{align}
  &\frac 1{Z(\M,\theta)}
    \int\frac{d\phi}{2\pi} \int \prod_k \frac{d\phi_k}{2\pi}\, 
    \delta\Big(\sum_k\phi_k-\theta +\phi\Big)f(\M,\vec\phi)\notag\\
  &\qquad\times\exp\Bigl[-2\left|\hu\sin\tfrac\phi2\right|+\sum_k\m_k \cos \phi_k \Bigr]\notag\\
  &= \frac 1{Z(\M,\theta)}
    \int \prod_k \frac{d\phi_k}{2\pi} f(\M,\vec\phi)\label{eq:Sasympt}\\
    &\qquad\times\exp\Bigl[-2\Bigl|\hu \sin\tfrac12\Bigl(\theta - \sum_k \phi_k\Bigr)\Bigr|+\sum_k \m_k\cos \phi_k \Bigr]\,.\notag
\end{align}
We now set the valence mass $\hu$ equal to one of the sea quark
masses, say $\m_1$. Using the trigonometric identity
$\cos\alpha-\cos\beta = 2 \sin[(\beta+\alpha)/2] \sin[(\beta-\alpha)/2]$, the exponent of the
last line in~\eqref{eq:Sasympt} can be written as
\begin{align}
  &-2\Big|\m_1\sin\tfrac{\theta-\sum_k\phi_k}2\Big|+\m_1\cos \phi_1
    -\m_1\cos\Big(\theta - \sum_{k\ge 2} \phi_k\Big)\notag\\
  &\quad+\m_1\cos\Big(\theta - \sum_{k\ge 2} \phi_k\Big)
    +\sum_{k\ge 2}\m_k \cos \phi_k\notag \\
  &=\!-2\Big|\m_1\sin\tfrac{\theta - \sum_k \phi_k}2\Big|
    \!+\!2\m_1\sin\tfrac{\theta+\phi_1-\sum_{k\ge2} \phi_k}2
    \sin\tfrac{\theta-\sum_k \phi_k}2\notag\\
  &\quad+\m_1 \cos\Big(\theta - \sum_{k\ge 2} \phi_k\Big)
    + \sum_{k\ge 2}\m_k \cos \phi_k\notag\\
  &\le\m_1 \cos\Big(\theta - \sum_{k\ge 2} \phi_k\Big)
    + \sum_{k\ge 2}\m_k \cos \phi_k\,,
    \label{est}
\end{align}
where in the last line we used the fact that the absolute value of the
second term is always smaller than the absolute value of the first
term. The RHS of this estimate is exactly the exponent of the
integrand for $Z(\M,\theta)$, cf.~\eqref{zmth}.
 
We have thus found that in Eq.~\eqref{sigsum0} the exponent of the
numerator is always smaller than or equal to the exponent of the
denominator. Therefore the sum of the zero-mode and quenched
contributions to the chiral condensate does not increase exponentially
for large rescaled masses $\m_k$.  The pre-exponential terms may have
(even strong) effects on the integral. They can be zero, can diverge
at the saddle points, or prevent us from reaching some of the saddle
points.  However, none of these effects can lead to an exponential
increase because the integral in \eqref{sigsum0} is well defined and
finite at fixed $\M$.

\begin{figure}[t!]
  \centerline{\includegraphics[width=.95\columnwidth]{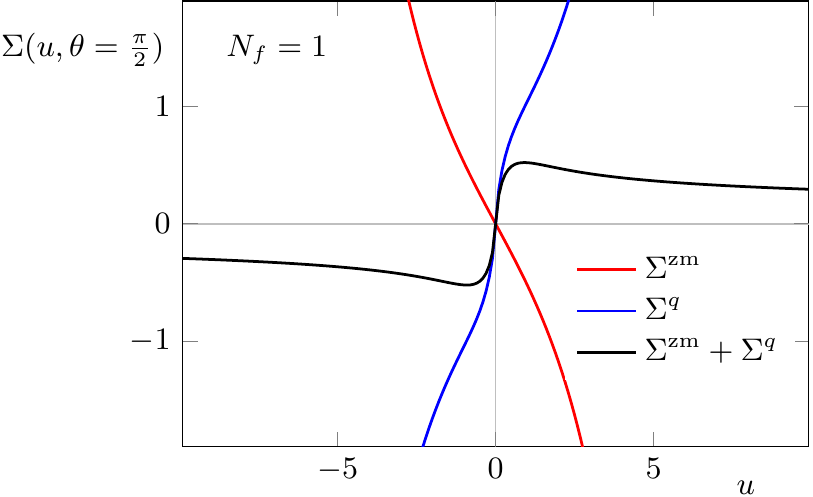}}
  \caption{Zero-mode (red) and quenched (blue) contribution to the
    chiral condensate for $N_f=1$ and $\theta = \pi/2$ as a function of the
    rescaled quark mass. Each contribution increases exponentially,
    but their sum (black) remains finite. }
\label{nf1can}
\end{figure}

\begin{figure}[b!]
   \centerline{\includegraphics[height=198mm]{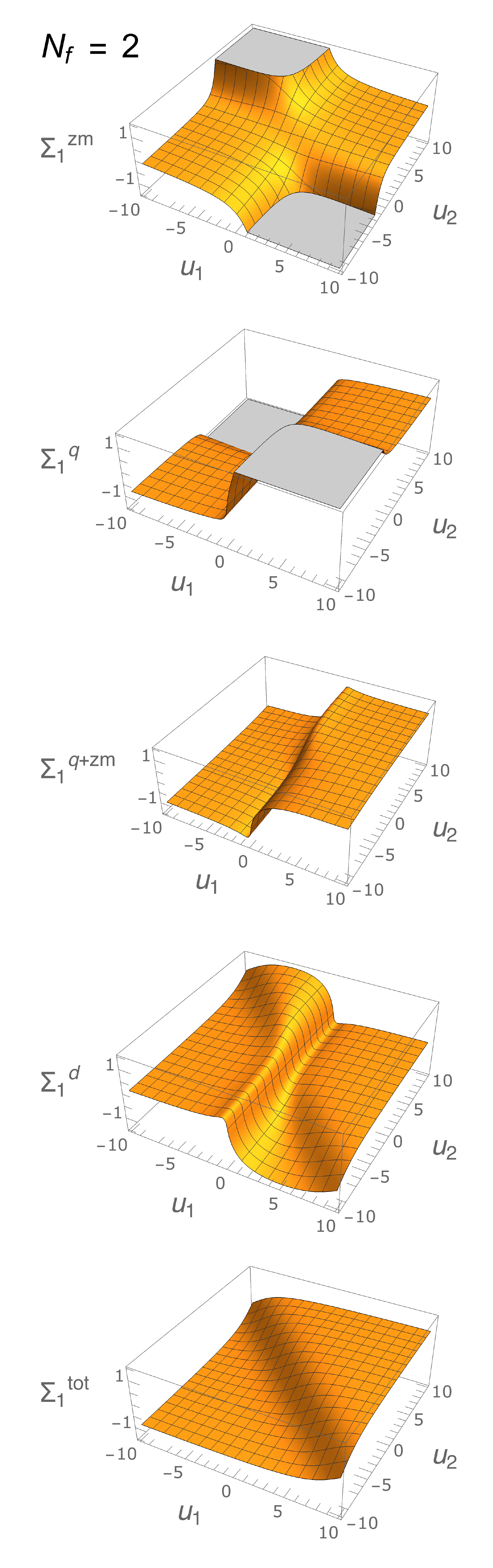}}
   \caption{Top to bottom: zero-mode ($\Sigma^\zm_1$) and quenched
     ($\Sigma^\q_1$) contributions to the chiral condensate, their sum
     ($\Sigma^\text{q+zm}_1=\Sigma^\q_1 +\Sigma^\zm_1$), and dynamical contribution
     ($\Sigma^{d}_1$) for $N_f=2$ and $\theta=0$ as a function of the rescaled
     quark masses $u_1$ and $u_2$.  The sum of all contributions,
     $\Sigma^\text{tot}_1 = \Sigma^\zm_1+\Sigma^\q_1+\Sigma^\d_1$ (bottom), shows a
     discontinuity at $u_1+u_2= 0$.  The subscript 1 of $\Sigma$ indicates
     that the condensate corresponds to the first quark.}
\label{nf2can}
\end{figure}

The exponential cancellation is illustrated in Fig.~\ref{nf1can} for
$N_f = 1$ at $\theta = \pi/2$ and in Fig. \ref{nf2can} for $N_f=2$ at
$\theta =0$, where both $\Sigma^\zm_1$ and $\Sigma^\q_1$ increase exponentially in
the quadrants where $\m_1\m_2 <0$. The sum of the two contributions
remains finite. Also shown in Fig.~\ref{nf2can} is the dynamical part
of the chiral condensate, see Eq.~\eqref{chiralcond-nzm-2f.b}.

Finally, we give a heuristic argument why the cancellation between the
zero-mode contribution and the quenched contribution to the chiral
condensate takes place. When we have $|\nu|$ zero modes, the nonzero
eigenvalues on the microscopic scale are, on average, shifted from the
origin by $|\nu|$, see Ref.~\cite[Sec. VIII.A]{Shifrin:2005cy}. To find
a definitive gap one needs to assume
$1\ll\nu\ll V\Lambda_\text{QCD}^4$.  Explicitly, one can easily show that the
asymptotic behavior of the microscopic quenched density is given by
\begin{align}
  \rho^{\q}_\nu(\x)  \overset{|\nu|\gg1}{\approx}  \frac{1}{ \pi} \frac{\sqrt{\x^2-\nu^2}}{|\x|}\theta(|x| -|\nu|)\,.  
\end{align}
The corresponding chiral condensate $\Sigma^{\q}_\nu(\hu)$ can be calculated
along Eq.~\eqref{eq:sigma}, i.e.,
\begin{align}
  \frac{\Sigma^{\q}_\nu(\hu)}{\Sigma}
  &\overset{|\nu|\gg1}{\approx}
    \frac{2\hu}{\pi}\int_{|\nu|}^{\infty}d\x\,\frac{\sqrt{\x^2-\nu^2}}{\x(\x^2+\hu^2)}\notag\\
  &\;\;\,=\frac{|\nu|}{\hu}\biggl(\sqrt{1+\frac{\hu^2}{\nu^2}}-1\biggr)\,.
\end{align}
For large masses at large, fixed topological index, the quenched
chiral condensate has the approximate form
\begin{align}\label{sig.q.approx}
  \frac{\Sigma^{\q}_\nu(\hu)}{\Sigma}\overset{|\hu|\gg|\nu|\gg1}{\approx}\sign(\hu)-\frac{|\nu|}{\hu}.
\end{align}
The second term shows that the nonzero-mode spectrum has been depleted
by $|\nu|$ modes so that the total number of eigenvalues does not depend
on $\nu$. The contribution of the zero modes is given by
$\Sigma_\nu^\zm(\hu)=|\nu|/\hu$.

At fixed (large) $\nu$ we thus find that for large $\hu$ the $1/\hu$
contribution to the quenched part of the chiral condensate exactly
cancels the contribution from the zero modes. At fixed $\theta$ the
contribution of the fermion determinant leads to exponentially large
terms (instead of $1/\hu$ terms) at large $\hu$, but as shown in
Sec.~\ref{section-4} we again have a cancellation between the
zero-mode and quenched parts.  This cancellation is deeply rooted in
topology and spectral flow, which guarantee that the total number of
eigenvalues around zero remains the same.  For chiral random matrix
theory, this can be shown at the technical level
\cite{Shifrin:2005cy}, but the argument is much more general: the
Dirac spectrum near zero is depleted by exactly the same number of
levels as we have zero modes.  One could argue that the spectral
density at fixed $\theta$-angle mainly involves very large $\nu$ so that in
the thermodynamic limit the spectrum acquires a gap at zero.  However,
since the topological susceptibility is finite in the thermodynamic
limit, the number of zero modes is of order $\sqrt V$, while the
eigenvalue density is of order $1/V$, resulting in a gap with a width
of order $1/\sqrt V$.

\section{One-Flavor QCD}
\label{sectio-5}

In this section we derive a number of explicit results for one-flavor
QCD. In the first subsection we compute the contributions to the
spectral density at fixed $\theta$-angle. In the second subsection we use
these results to compute the dynamical contribution to the chiral
condensate and show that a mass-independent total chiral condensate is
obtained.
      
\subsection{\boldmath One-flavor Dirac spectrum at fixed $\theta$-angle}
\label{sec:one-flavor}

The one-flavor partition function of QCD in the $\epsilon$-domain at fixed
$\theta$ is given by
\begin{align}
  Z(\m,\theta) = e^{\m \cos \theta}\,,
\end{align}
see Eq.~\eqref{one}.  At fixed $\nu$ we therefore have from
Eq.~\eqref{eq:inverse}
\begin{align}
  Z_\nu(\m) = \int_{-\pi}^\pi \frac{d \theta}{2\pi}\, e^{-i\nu\theta+\m \cos \theta} = I_\nu(\m)\,.
\end{align}
The spectral density at fixed $\theta$ is given by
\begin{align}
  \label{eq:rho_theta}
  \rho(\x,\m,\theta) = \frac 1{e^{\m\cos\theta}}\sum_\nu e^{i\nu\theta}
  I_\nu(\m) \rho_\nu(\x,\m)\,.  
\end{align}
In the previous section we already obtained the zero-mode part and the
quenched part of the spectral density, see Eqs.~\eqref{eq:rhozm}
and~\eqref{eq:rhoqtheta}, respectively. Explicitly, the zero-mode part
reads
\begin{align}
  &\rho^\zm(\x,\m,\theta)=\delta(x)\dashint_{-\pi}^\pi\frac{d\varphi}{2\pi}\,
  \frac{1-e^{\m[\cos(\theta-\varphi)-\cos\theta]}}{2\sin^2\frac\phi2}\notag\\
  &=\delta(x)\m\dashint_{-\pi}^\pi\frac{d\varphi}{2\pi}\,
    e^{\m[\cos(\theta-\varphi)-\cos\theta]}\sin(\phi-\theta)\cot\tfrac\phi2\,,  
    \label{eq:rhozm1}
\end{align}
where the second line follows after partial integration.  The quenched
part is equal to
\begin{equation}
  \rho^\q(\x,\m,\theta)=\int_{-\pi}^\pi\frac{d\phi}{2\pi}
  \frac{J_1(2|\x|\sin\frac\phi2)}{2\sin\frac\phi2}
  e^{\m[\cos(\theta-\phi)-\cos\theta]}\,.
\end{equation}
The dynamical part at fixed $\nu$ is given by
\cite{Damgaard:1997ye,Wilke:1997gf}
\begin{align}\label{sum-rhod}
  \rho^\d_\nu(\x,\m)=\frac {-|\x|}{\x^2+\m^2}
  \Big [\x J_\nu(\x) J_{\nu+1}(\x)+\m \frac
  {I_{\nu+1}(\m)}{I_\nu(\m)}J_\nu^2(\x)\Big]. 
\end{align}
After performing the sums over $\nu$ in Eq.~\eqref{eq:rhotheta} with the
help of Eq.~\eqref{a1.10} we obtain for the dynamical part at fixed
$\theta$
\begin{align}  \label{rhodyn}
  &\rho^\d(\x,\m,\theta) = -\frac {|\x|}{\x^2+\m^2}
  \int_{-\pi}^\pi\frac{d\phi}{2 \pi}\, e^{\m[\cos(\theta-\phi)-\cos\theta]}\\
  &\qquad\times\left[\x\sin\tfrac\phi2 J_1(2\x \sin\tfrac\phi2)
    +\m \cos(\theta-\phi)J_0(2\x \sin \tfrac\phi2)\right]. \notag
\end{align}
Note that the imaginary part resulting from \eqref{a1.10} is the
integral of a total derivative and therefore vanishes. Moreover, the
result has to be real since the expression~\eqref{sum-rhod} is
invariant under $\nu\leftrightarrow-\nu$, which can be shown using the recursion
relations \eqref{J.rec} and \eqref{I.rec} of Bessel functions.

Adding the dynamical part to the quenched part and performing some
simplifications, the nonzero-mode part of the density at fixed
$\theta$ is given by
\begin{align}
\label{density-non-Nf1-theta}
  &\rho^\nzm(\x,\m,\theta)=\frac{|\x|}{\x^2+\m^2}
    \int_{-\pi}^\pi\frac{d\phi}{2\pi}\,e^{\m[\cos(\theta-\phi)-\cos\theta]}\\
  &\Bigl[(\m^2\!+\!\x^2\cos\phi)
    \frac{J_1\big(2\x\sin\frac\phi2\big)}{2\x\sin\frac\phi2}
    -\m\cos(\theta\!-\!\phi)J_{0}\big(2\x\sin\tfrac{\phi}{2}\big)\Bigl],\notag
\end{align}
which for $\theta = 0$ was already obtained in
\cite{Verbaarschot:2014qka,Verbaarschot:2014upa}.

\begin{figure}[t!]
  \centerline{\includegraphics[width=0.8\columnwidth]{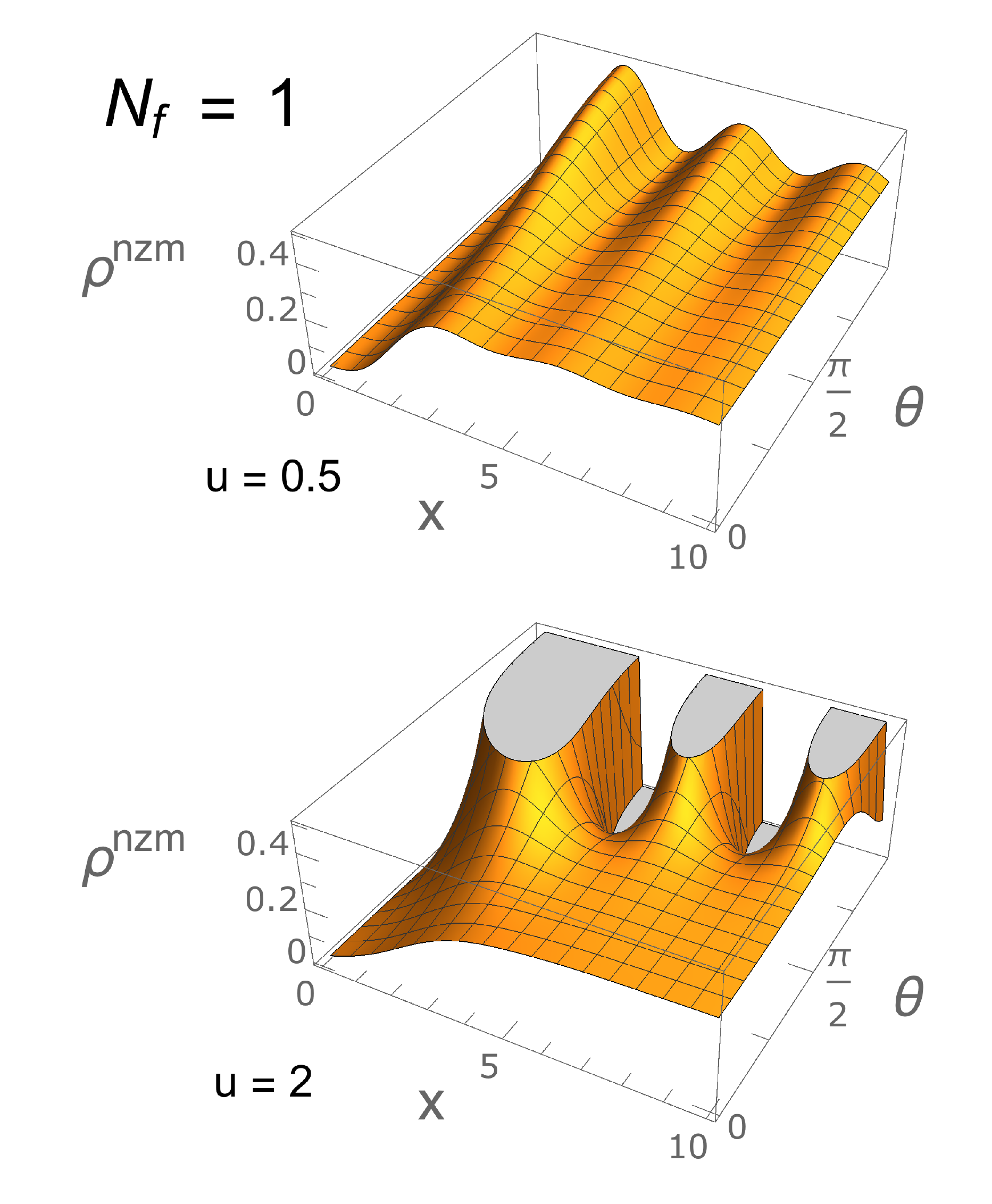}}
  \caption{Microscopic spectral density~\eqref{density-non-Nf1-theta}
    of the one-flavor theory as a function of the $\theta$-angle for
    the two rescaled masses $\m=0.5$ and $\m=2$. The oscillations
    become stronger with increasing $\theta$ and $\m$ and eventually
    yield an exponentially large spectral density. The reason for this
    divergence is the non-positive statistical weight due to the
    $\theta$-angle. }
  \label{fig:rhoNf1}
\end{figure}

In Fig.~\ref{fig:rhoNf1} we show the spectral density of the nonzero
modes at fixed $\theta$ for rescaled quark masses $\m=0.5$ and $\m=2$. At
nonzero $\theta$ the amplitude of the oscillations increases exponentially
with $\m$. The only exception is $\theta =\pi $, where the spectral density
is well behaved for large negative mass but increases exponentially
with positive mass $\m$, see the discussion prior to
Eq.~\eqref{eq:Sigmam}.

For large mass $\m$, a saddle-point analysis of
Eq.~\eqref{density-non-Nf1-theta} shows that the spectral density
behaves as
\begin{align}
  \rho^\nzm(\x,\m,\theta)\overset{|\m|\gg1}{\approx}
  e^{|\m|-\m\cos\theta}\,\frac{J_1(2|\x|\sin\tfrac{\theta_0}2)}
  {\sqrt{8\pi|\m|}\sin \tfrac{\theta_0}2}
\end{align}
with $\theta_0 =\theta+[1-\sign(\m)]\pi/2$.  When the exponent vanishes, i.e., for
$\theta = 0$ with positive mass or $\theta = \pi$ with negative mass, the
asymptotic expansion is still valid, resulting in
\begin{equation}
  \rho^\nzm(\x,\m,0)\overset{|\m|\gg1}{\approx}\frac{|\x|}{\sqrt{8\pi|\m|}}\,.
\end{equation}

\subsection{Chiral condensate}
\label{sec:cc1}

In this subsection we use Eq.~\eqref{eq:sigma} to compute the chiral
condensate for one flavor and show that it is mass independent. Here
and in Sec.~\ref{sec:Nf1-limit}, the valence quark mass $\hu$ is set
equal to the sea quark mass $\m$.

The sum of the zero-mode and quenched contributions was already
computed in Eq.~\eqref{sigsum0}. It is given by
\begin{align}  \label{sigsum}
  &\frac{\Sigma^\zm(\m,\theta)}\Sigma+\frac{\Sigma^\q(\m,\theta)}\Sigma\\
  &=\dashint_{-\pi}^\pi\frac{d\phi}{2\pi}
  \left[ \frac 1{2u \sin^2\frac\phi 2}-
    \frac{\Kuphi1}{\sign(\m)|\sin(\frac \phi 2)|}\frac {e^{\m\cos(\theta-\phi)}}
      {e^{\m \cos\theta}} \right ]
  .\notag
\end{align}
The contribution from the dynamical part of the density follows from
Eqs.~\eqref{eq:sigma} and \eqref{rhodyn} using Eqs.~\eqref{eq:J12} and
\eqref{eq:J01},
\begin{align}
  \frac{\Sigma^\d(\m,\theta)}\Sigma &= -2\int_{-\pi}^\pi\frac{d\phi}{2 \pi}\,
    \frac{e^{\m\cos(\theta-\phi)}}{e^{\m\cos\theta}}
    \Bigl[\m\sin^2\tfrac\phi2\Kuphi0 \notag\\
  &\quad+ |\m\sin\tfrac\phi2| \cos(\theta-\phi)\Kuphi1\Bigr].
  \label{sigma-dyn-therm-f1}
\end{align}
Since the leading asymptotic behavior of the $K_\nu$ Bessel functions,
see Eq.~\eqref{eq:Kasympt}, does not depend on the index we can use
the arguments of Sec.~\ref{section-4} to show that there are no
exponentially increasing contributions from the dynamical part. It is
noteworthy that this argument holds for an arbitrary number of flavors
$N_f$. Adding the last two equations we obtain the total chiral
condensate
\begin{align}
\frac{\Sigma(\m,\theta)}\Sigma
&=\dashint_{-\pi}^\pi\frac{d\phi}{2\pi}\,\biggl\{ \frac 1{2\m\sin^2\frac\phi2}
-\frac{e^{\m\cos(\theta-\phi)}}{e^{\m\cos\theta}}\notag\\
&\quad\times\Bigl [\m \Kuphi2 -\m\cos\phi\, \Kuphi0\notag\\
&\quad+2|\m\sin\tfrac{\phi}{2}|\cos(\theta\!-\!\phi) \Kuphi1 \Bigr]\biggr\}\,,
\end{align}
where we used~\eqref{Bessel-K-rec.a}. Using the recursion relations
\eqref{Bessel-K-rec.a} and~\eqref{Bessel-K-rec.b}, the Bessel function
$K_2$ can be rewritten as
\begin{align}
   &\Kuphi2=-\cos\phi \Kuphi0\notag\\
   &\qquad-\frac{2}{|\m|}\sign\left(\sin\tfrac{\phi}{2}\right)
     \frac{\partial}{\partial\phi}\left[\cos\tfrac{\phi}{2} \Kuphi1\right].
  \label{K2-rew}
\end{align}
Then the derivative can be integrated by parts, where the boundary
terms of the principal-value integral vanish. Hence we have
\begin{align}
  &\frac{\Sigma(\m,\theta)}{\Sigma}
  =2\dashint_{-\pi}^\pi\frac{d\phi}{2\pi}\,\frac{e^{\m\cos(\theta-\phi)}}{e^{\m\cos\theta}}
  \Bigl [\m\cos\phi \Kuphi0  \notag\\
  &\qquad-|\m|\sign\left(\sin\tfrac{\phi}{2}\right)\sin(\theta-\tfrac{\phi}{2})\Kuphi1\Bigr].
\label{final-sig1}  
\end{align}
To show that the principal-value integral is indeed equal to
$\cos\theta$, cf.~Eq.~\eqref{eq:Sigmaonef}, we first note that for
$\m=0$ we indeed obtain
\begin{equation}
  \frac{\Sigma(\m=0,\theta)}\Sigma
  =-\dashint_{-\pi}^\pi\frac{d\phi}{2\pi}\,\frac{\sin(\theta-\frac\phi2)}{\sin\frac\phi2}
  =\cos\theta\,,
\end{equation}
where we used the asymptotics $K_1(|x|)\approx1/|x|$ for $|x|\ll1$.  In the
second step we take the mass derivative of the chiral condensate,
which is an integral of a total derivative and thus vanishes,
\begin{widetext}
\begin{align}
  \frac d{d\m}\Sigma(\m,\theta) &=2\dashint_{-\pi}^\pi\frac{d\phi}{2\pi}\,\frac{e^{\m\cos(\theta-\phi)}}{e^{\m\cos\theta}}\Bigl [\cos\phi \Kuphi0-2|\m\sin\tfrac{\phi}{2}|\cos\phi \Kuphi1+2\m\sin\tfrac{\phi}{2}\sin(\theta-\tfrac{\phi}{2})\notag\\
  &\quad\times\Kuphi0+2\sin(\theta-\tfrac{\phi}{2})\sin\tfrac{\phi}{2}\Bigl(\m\cos\phi \Kuphi0-|\m|\sign\left(\sin\tfrac{\phi}{2}\right)\sin(\theta-\tfrac{\phi}{2})\Kuphi1\Bigr)\Bigr]\notag\\
   &=2\int_{-\pi}^\pi \frac{d\phi}{2\pi}\, \frac d{d\phi}
    \frac{e^{\m\cos(\theta-\phi)}}{e^{\m\cos\theta}}\Bigl[\sin\phi\,\Kuphi0-2\sign(\m) \sin\theta |\sin \tfrac \phi 2| \Kuphi1\Bigr]=0\,.
\end{align}
\end{widetext}
Therefore we conclude that $\Sigma(\m,\theta)$ does not depend on $u$ so that
\begin{align}
\frac{  \Sigma(\m,\theta)}\Sigma = \cos\theta\,.
\end{align}

The derivation above also shows that the nonzero-mode contribution to
the chiral condensate can be simplified to
\begin{align}
  &\frac{\Sigma^\nzm(\m,\theta)}\Sigma = \cos \theta - \frac{\Sigma^\zm(\m,\theta)}\Sigma\notag\\
  &= \cos \theta -  \dashint_{-\pi}^\pi \frac {d\phi}{2\pi}\,
    \frac{e^{\m\cos(\theta-\phi)}}{e^{\m\cos\theta}}  \sin(\phi-\theta) \cot \tfrac \phi 2\,,
    \label{eq:sigmanzm-Nf1}
\end{align}
which follows from Eq.~\eqref{eq:rhozm1}.

\begin{figure}[h]
  \centerline{\includegraphics[width=\columnwidth]{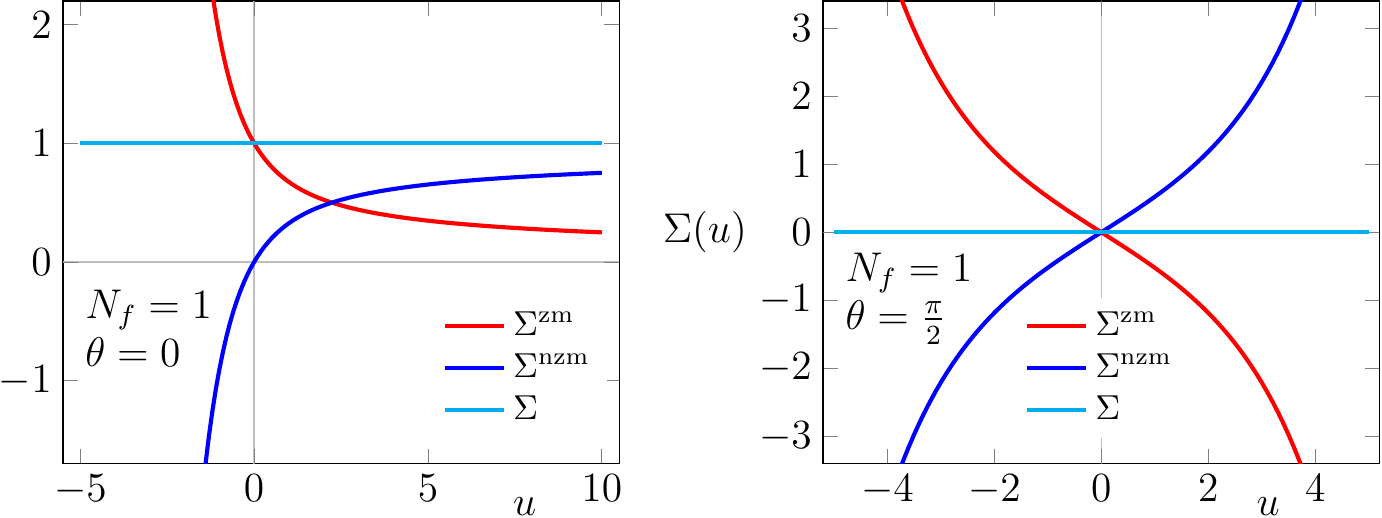}}
  \caption{Chiral condensate for one-flavor QCD and its splitting into
    the contributions from zero and nonzero modes.  Results are given
    for $\theta=0$ (left) and $\theta=\pi/2$ (right). While for positive mass and
    vanishing $\theta$-angle the contributions behave algebraically they
    increase exponentially with the volume otherwise, regardless of
    whether the sign/phase comes from the mass or a non-vanishing
    $\theta$-angle. }
  \label{fig:chiralNf1}
\end{figure}

The zero-mode and nonzero-mode contributions to the chiral condensate
are plotted as a function of $\m$ in Fig.~\ref{fig:chiralNf1} for
$\theta=0$ and $\pi/2$.  It becomes obvious that when a nontrivial phase is
present the contributions of both the zero and nonzero modes grow
exponentially with the volume (included in $\m$), but that they add up
to a finite result that gives a mass-independent chiral condensate.

\subsection{Thermodynamic limit}\label{sec:Nf1-limit}

We first consider the case when there is no sign problem, i.e.,
$\cos\theta=\sign\m$.  For this case the thermodynamic limit
$\m\to\infty$ can readily be derived for the zero-mode contribution
\cite[Eq.~(7.3)]{Leutwyler:1992yt},
\begin{align}
  \label{density-zero-Nf1-theta-special-limit}
  \rho^\zm(\x,\m>0,\theta=0)&= \delta(\x)|\m|e^{-|\m|}[I_0(|\m|) +I_1(|\m|) ] \notag\\
  &\overset{\makebox[0mm]{\scriptsize$|\m|\gg1$}}{\approx}\quad\delta(\x)\sqrt{\frac{2 |\m|}{\pi}}\,,\notag \\
  \frac{\Sigma^\zm(\m>0,\theta=0)}\Sigma &\overset{|\m|\gg1}{\approx}\sqrt{\frac{2}{\pi|\m|}}\,.
\end{align}
Therefore the contribution of the zero modes to the chiral condensate
behaves like $1/\sqrt{|\m|}$. This can be seen in the left plot of
Fig.~\ref{fig:chiralNf1} for positive masses. For $u < 0$ and
$\theta = \pi$ we find the same asymptotical spectral density as in
Eq.~\eqref{density-zero-Nf1-theta-special-limit}, but the chiral
condensate differs by a minus sign.

We now turn to the general case with a sign problem. In this case we
have to perform a saddle-point analysis for the integral determining
the level density and the chiral condensate. This is most easily done
in Eq.~\eqref{eq:rhozm1}.  Expanding $\phi$ about the point where the
exponent is maximized, namely $\cos(\phi-\theta)=\sign\m$, we obtain
\begin{align}
  \rho^\zm(\x,\m,\theta)&=\delta(\x)\m\Sigma^\zm(\m,\theta)/\Sigma\,,\notag\\
  \frac{\Sigma^\zm(\m,\theta)}\Sigma&\overset{|\m|\gg1}{\approx}
  -\frac{e^{|\m|(1-\sign(\m)\cos\theta)}}{\sqrt{2\pi|\m|^3}(\sign(\m)-\cos\theta)}\,.
  \label{density-zero-Nf1-theta-limit}
\end{align}
Here, we notice the main difference between the sign-quenched
result~\eqref{density-zero-Nf1-theta-special-limit} and the result
with a sign problem. The contribution of the zero modes to the chiral
condensate vanishes as $1/\sqrt{|\m|}$ in the sign-quenched case,
while it diverges exponentially regardless of how small the angle
$\theta+\arg\m$ is. In particular, the divergence is strongest when
$\theta+\arg\m=\pm\pi$, which reflects the sign problem observed in the
difference of the free energies~\eqref{free-energy-Nf1}. Additionally,
the sign of the contribution of the zero modes to the chiral
condensate changes with the sign of the quark mass. Its behavior for
$\theta = 0$ and $\theta = \pi/2$ is shown in Fig.~\ref{fig:chiralNf1}.

Whether or not a sign problem is present, the contribution of the
nonzero modes to the chiral condensate is equal to the difference
$\Sigma\cos\theta - \Sigma^\zm$.

We now consider the thermodynamic limit of the dynamical part
$\Sigma^\d$, see Eq.~\eqref{sigma-dyn-therm-f1}, which stays finite for
large masses.  Its asymptotic behavior is worked out in
App.~\ref{app:asympt}, resulting in
\begin{align}\label{sig-d-f1-lim}
  \frac{\Sigma^\d(\m,\theta)}\Sigma\overset{|\m|\gg1}{\approx}
  \begin{cases}
    -\frac{1}{2\m\cos^2\theta}\,, & \m\cos\theta>0 \,,\\
    2\cos\theta\,, & \m\cos\theta<0 \,,\\
    -\frac{\sign(u)\Gamma(5/6)}{\sqrt\pi|u|^{1/3}}\,, & u\cos\theta=0\,,
  \end{cases}
\end{align}
where $\Gamma(5/6)$ in the last case denotes the $\Gamma$-function.

Finally, to identify the thermodynamic limit of the quenched
contribution we can combine the relation
$\Sigma(\m,\theta)=\Sigma^\zm(\m,\theta)+\Sigma^q(\m,\theta)+\Sigma^d(\m,\theta)=\Sigma\cos\theta$ with the results
\eqref{density-zero-Nf1-theta-special-limit} through
\eqref{sig-d-f1-lim}.

\section{Two-Flavor QCD}
\label{section-6}

In the first subsection we compute the two-flavor spectral density at
fixed $\theta$-angle, which is used in the second subsection to compute the
various contributions to the chiral condensate.

\subsection{\boldmath Two-flavor Dirac spectrum at fixed
  $\theta$-angle}\label{sec:two-flavor}

The spectral density at fixed $\theta$-angle is again given by
Eqs.~\eqref{eq:rhotheta} and \eqref{eq:Ptheta}. The two-flavor
partition function at fixed $\theta$-angle has already been given in
\eqref{eq:Z2}. At fixed topological charge $\nu$ it can be written as
\cite{Jackson:1996jb}
\begin{equation}\label{part-Nf2-nu}
  Z_\nu(\M)=2\frac{\m_1I_{\nu+1}(\m_1)I_{\nu}(\m_2)
  -\m_2I_{\nu+1}(\m_2)I_\nu(\m_1)}{\m_1^2-\m_2^2}\,.
\end{equation}
The zero-mode contribution to the spectral density is determined by
Eq.~\eqref{eq:rhozm} with $Z(\M,\theta-\varphi)$ as in Eq.~\eqref{eq:Z2}.

Next we evaluate the nonzero-mode contribution to the spectral
density.  At fixed $\nu$ the spectral density is given by
\cite{Damgaard:1997ye,Wilke:1997gf}
\begin{align}
  \label{rho-two}
  &\rho_\nu(\x,\M)=
  \frac {|\x|}{Z_\nu(\M) (\x^2+\m_1^2)(\x^2+\m_2^2)(\m_2^2-\m_1^2)}
\\&\times
  \begin{vmatrix}
   J_{\nu-1}(\x)/\x & J_\nu(\x) & I_\nu(-\m_1)& I_\nu(-\m_2) \\
   J_{\nu}(\x) & \x J_{\nu+1}(\x) &\m_1 I_{\nu+1}(-\m_1)&\m_2 I_{\nu+1}(-\m_2) \\
  \x J_{\nu+1}(\x) & \x^{2} J_{\nu+2}(\x) &\m_1^{2} I_{\nu+2}(-\m_1) &\m_2^{2} I_{\nu+2}(-\m_2) \\\x^{2} J_{\nu+2}(\x) & \x^{3} J_{\nu+3}(\x) &\m_1^{3} I_{\nu+3}(-\m_1)&\m_2^{3} I_{\nu+3}(-\m_2)
  \end{vmatrix},\notag
\end{align}
where $|\cdot|$ denotes the determinant of the matrix.  The determinant in
Eq.~\eqref{rho-two} can be rewritten as
\begin{widetext}
  \begin{align}
    \begin{vmatrix}
      \x^{-1} J_{\nu-1}(\x) & J_\nu(\x) & I_\nu(-\m_1)& I_\nu(-\m_2) \\
      J_{\nu}(\x) & \x J_{\nu+1}(\x) &\m_1 I_{\nu+1}(-\m_1)&\m_2 I_{\nu+1}(-\m_2) \\
      -2 J_{\nu}(\x) & 0 &(\m_1^{2}+\x^2) I_{\nu}(-\m_1) & (\m_2^{2}+\x^2) I_{\nu}(-\m_2) \\
      -2\x J_{\nu+1}(\x) & 0 & \m_1(\m_1^2+\x^2) I_{\nu+1}(-\m_1)& \m_2(\m_2^2+\x^2) I_{\nu+1}(-\m_2)
    \end{vmatrix}
  \end{align}
\end{widetext}
by employing the recurrence relations~\eqref{J.rec} and~\eqref{I.rec}
and the properties of the determinant.  In this form the terms
involving the upper left $2\times 2$ block yield the quenched level
density~\eqref{eq:rhoq}. The remaining terms represent the dynamical
part and can be simplified to
\begin{align} \label{level-quenched-f2}
&\rho^\d_\nu(x,\M) =-\frac{2|\x|}{Z_\nu(\M) (\x^2+\m_1^2)(\x^2+\m_2^2)}\\
&\!\times\!\bigl\{\x J_\nu(\x)J_{\nu+1}(\x)[\m_1I_{\nu+1}(\m_1)I_\nu(\m_2)\!+\!\m_2I_{\nu+1}(\m_2)I_\nu(\m_1)]
\notag\\
&\!+\x^2J_{\nu+1}^2(\x)I_\nu(\m_1)I_\nu(\m_2)\!+\!\m_1\m_2J_{\nu}^2(\x)I_{\nu+1}(\m_1)I_{\nu+1}(\m_2)\bigr\}\notag
\end{align}
by employing the symmetry $I_\nu(-\m)=(-1)^\nu I_\nu(\m)$. Note that the
dynamical part $\rho_\nu^\d$ is again symmetric in $\nu\to-\nu$ because
$\rho_\nu$ and $\rho_\nu^\q$ are, so that the sum including the phases
$e^{i\nu \theta}$ yields a real function.

Now we sum over $\nu$ as shown in~\eqref{eq:rhotheta}. The quenched
contribution $\rho^\q$ was already obtained in~\eqref{eq:rhoqtheta} with
the two-flavor partition function $Z(\m_1,\m_2,\theta)$, see
Eq.~\eqref{eq:Z2}.  The sum for the dynamical part is more involved,
and the required sums over products of four Bessel functions are
worked out in Eq.~\eqref{a1.11}.  Adding the quenched contribution,
this results in the total spectral density of the nonzero modes,
\begin{align}
  &\rho^\nzm(\x,\M,\theta) = \frac{|\x|}{Z(\M,\theta)}
    \int_{-\pi}^\pi\!\frac{d\phi}{2\pi}\biggl[
    \frac{ J_1(2\x\sin\frac \phi 2)}{2\x\sin\frac \phi 2}Z(\M,\theta-\phi)\notag\\
  &+\frac{\x (2\m_1\m_2e ^{i( \phi -\theta)}+\m_2^2+\m_1^2  )
    J_1(2\x\sin\frac \phi 2)}{(\x^2+\m_1^2)(\x^2+\m_2^2)ie^{i\frac \phi 2}}
    Z(\M,\theta-\phi)\notag\\ 
  &-\frac{2(\m_1\m_2 e^{i(\phi-\theta )} +\x^2 e^{-i\phi}  ) J_0(2\x\sin\frac \phi 2) }{(\x^2+\m_1^2)(\x^2+\m_2^2)}
    \zeta_0(\M,\theta-\phi)\biggr]
\end{align}
with
\begin{align}
  \zeta_0(\M,\alpha)=I_0\Bigl(\sqrt{\m_1^2+\m_2^2 +2\m_1\m_2\cos\alpha}\Bigr)\,.
\end{align}
We recognize the first term as the contribution of the quenched part
of the spectral density discussed in Sec.~\ref{section-4}.  The
imaginary part of the integral vanishes, which follows from the fact
that the imaginary part resulting from applying Eq.~\eqref{a1.11}
yields a total derivative.
    
The level density of the nonzero-modes, $\rho^\nzm=\rho^\q+\rho^\d$, looks
quite complicated. It becomes more presentable when both masses are
equal and the $\theta$-angle vanishes,
\begin{align}
  &\rho^{\nzm}(\x,u,u,\theta=0)=\frac{|\x\m|}{I_1(2|\m|)}
    \int_{-\pi}^\pi\frac{d\phi}{\pi}\biggl[
    \frac{J_1\left(2\x\sin \frac \phi 2\right)}{2\x\sin \frac \phi 2 } \notag\\
  &\times  \frac{I_1\left(2\m\cos \frac \phi 2 \right)}{2\m\cos \frac \phi 2}
 -\cos\phi \frac{J_0(2\x\sin\frac{\phi}{2})I_0(2\m\cos\frac{\phi}{2})}{\x^2+\m^2} \biggl]. \label{density-non-Nf2-theta-special.2}
\end{align}
\begin{figure}[b]
  \centerline{\includegraphics[width=0.8\columnwidth]{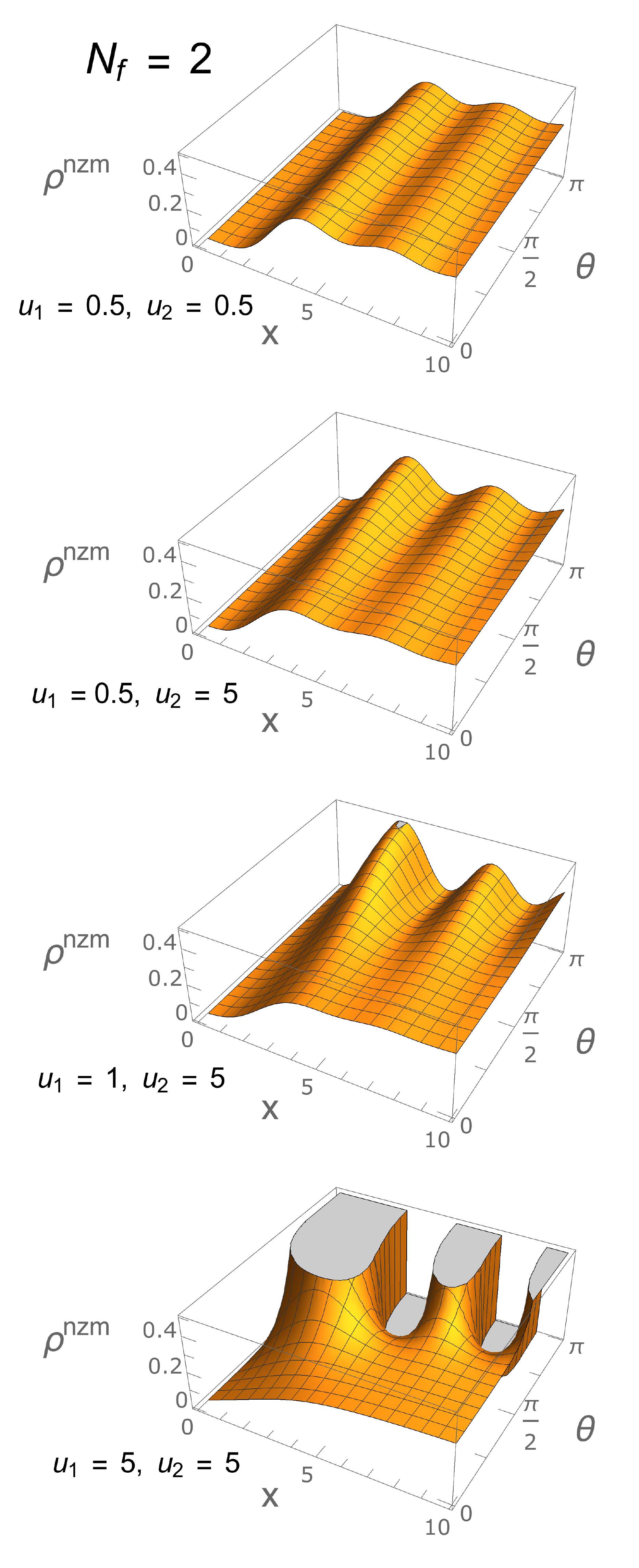}}
  \caption{Microscopic level density for two flavors as a function of
    the eigenvalue position $\x$ and of the $\theta$-angle for various
    quark masses $\m_1$ and $\m_2$. When increasing the quark masses
    the oscillations become so strong that the level density loses its
    positivity. }
  \label{fig:rhoNf2}
\end{figure}
The behavior of the level density $\rho^{\nzm}$ is shown in
Fig.~\ref{fig:rhoNf2} for various masses. As in the one-flavor case,
the amplitude of the level density at nonzero $\theta$-angle increases
exponentially with the volume, and its oscillations have a period of
$O(1/V)$. At nonzero $\theta$-angle, oscillations of this type can shift
the original discontinuity of the chiral condensate at $m=0$ for fixed
$\nu$.

\begin{figure}[b]
  \centerline{\includegraphics[width=0.8\columnwidth]{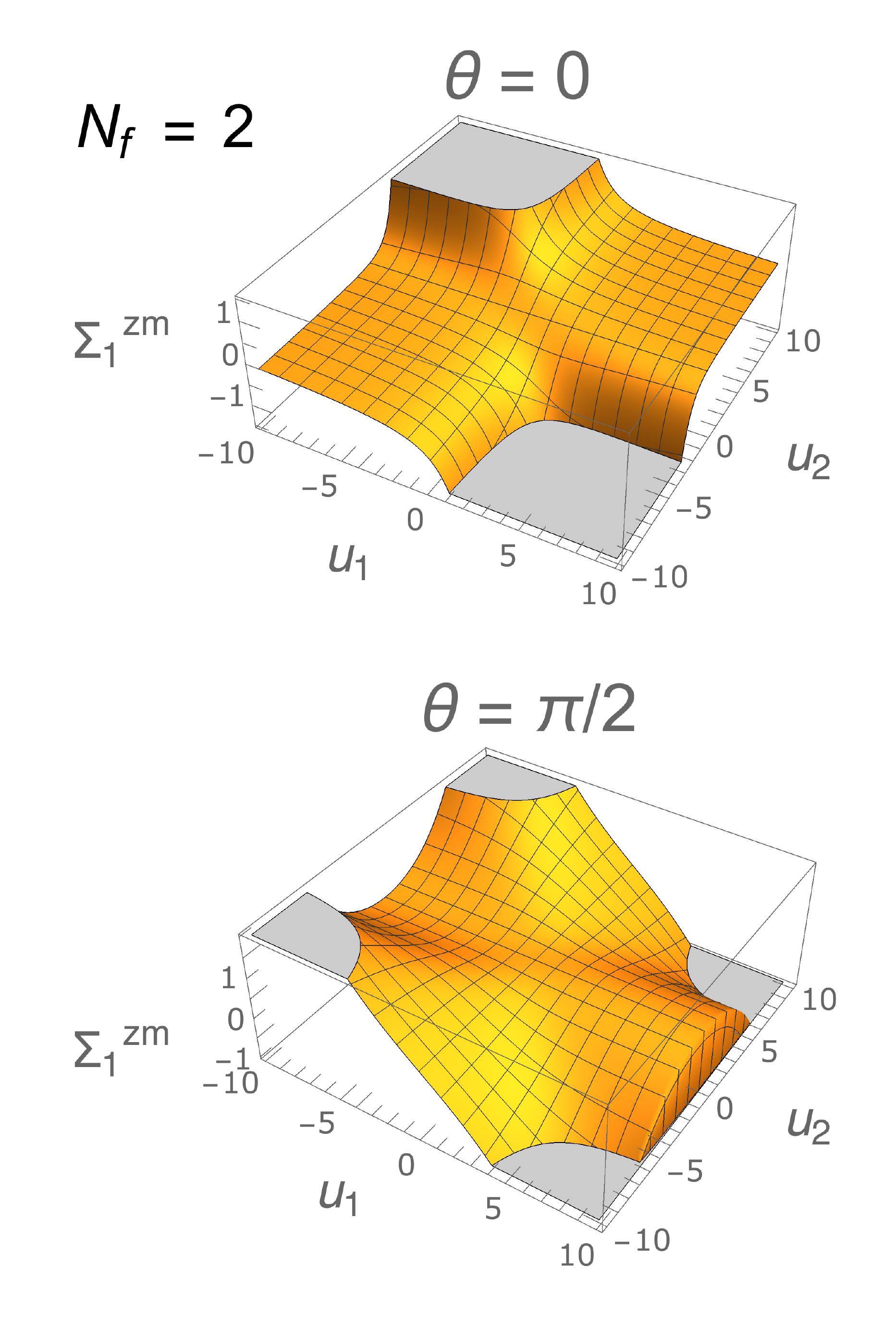}}
  \caption{Zero-mode contribution to the chiral condensate for two
    flavors as a function of the two quark masses $\m_1$ and
    $\m_2$. For vanishing $\theta$-angle we notice two quadrants where this
    contribution does not grow exponentially in the masses. For
    $\theta\in(0,\pi)$ we have only two lines, given by $\m_1\m_2=0$, where this
    contribution remains finite in the limit of large masses.}
  \label{fig:chiralNf2b}
\end{figure}

For large mass $\m_2$ we regain the level density
$\rho^\nzm_{N_f=1}(\x,\m_1,\theta)$. To obtain this limit we have to
approximate the modified Bessel functions by
\begin{align}\label{eq:Iasympt}
  I_\nu\Bigl(\sqrt{\m_1^2+\m_2^2+2\m_1\m_2\cos\phi}\Bigr)
  \approx\frac{e^{|\m_2|+\sign(\m_2)\m_1\cos\phi}}{\sqrt{2\pi |\m_2|}}\,.
\end{align}
In contrast, when taking the limit $\m_2\to 0$, only the sector of zero
topological charge contributes to the partition function, resulting in
the spectral density $\rho^\nzm_{\nu=0}(\x,\m_1,0)$. Because of
flavor-topology duality \cite{Verbaarschot:1997bf} for massless quarks
in the microscopic limit (which is most easily understood in terms of
the joint eigenvalue distribution of the chiral random matrix theory),
this can be written as $\rho^\nzm_{\nu=N_{f}=1}(\x,\m=\m_1)$.

\subsection{Chiral Condensate}

In this subsection we evaluate the chiral condensate from the spectral
density using Eq.~\eqref{eq:sigma}. We only consider the chiral
condensate of the first quark (which we denote by $\Sigma_1$), i.e., we set
$\hu=\m_1$. The chiral condensate $\Sigma_2$ of the second quark can then
be obtained by interchanging $\m_1\leftrightarrow \m_2$.

The zero-mode and quenched contributions to the chiral condensate are
obtained by substituting the two-flavor result \eqref{eq:Z2} for the
partition function in Eqs.~\eqref{eq:sigmazm}, \eqref{sigqgen} and
\eqref{sigsum0}. We do not repeat the corresponding expressions here.
As in the one-flavor case, $\Sigma^{\zm}_1(\M,\theta)$ grows exponentially in
the quark masses for $\theta \ne 0$ or for $\theta = 0$ and
$\m_1\m_2 <0$. This can be seen nicely in
Fig.~\ref{fig:chiralNf2b}. When the sign problem is absent,
$\Sigma^{\zm}_1$ remains bounded. For vanishing $\theta$-angle we can use
$-1/2\sin^2\frac\varphi2=\partial_\varphi\cot\frac\varphi2$ and integrate
Eq.~\eqref{eq:sigmazm} by parts such that the integral loses its
singularity and becomes an ordinary integral,
\begin{align}\label{chiralcond-zero-Nf2-theta-special}
  \frac{\Sigma^{\zm}_1(\M,0)}\Sigma&=\frac{2\m_2|\m_1+\m_2|}{I_1(|\m_1+\m_2|)}
  \int_{-\pi}^\pi\frac{d\phi}{2\pi}\,\cos^2\tfrac\phi2\notag\\
  &\qquad\times\frac{I_2(\sqrt{\m_1^2+\m_2^2+2\m_1\m_2\cos\phi})}{\m_1^2+\m_2^2+2\m_1\m_2\cos\phi}\,.
\end{align}
We employed Eq.~\eqref{I.rec.diff} to obtain this result. A similar
result can be derived for $\theta=\pi$.  When the masses are equal, the
integral can be expressed in terms of a hypergeometric function.

The calculation of the dynamical contribution to the chiral condensate
is performed in App.~\ref{sec:4} and yields
\begin{widetext}
  \begin{align}
    &\frac{\Sigma_1^\d(\vec \m,\theta)}\Sigma=u_1\int_{-\pi}^\pi\frac{d\phi}{\pi}
      \biggl\{ \left[2 \m_1\m_2 \sin \tfrac \phi 2 \sin(\tfrac \phi 2- \theta)
      - (\m_1^2+\m_2^2)\sin^2\tfrac  \phi 2 \right]
      \biggl[\frac{K_0(2|\m_1\sin\tfrac \phi 2|)}{\m_2^2-\m_1^2}\notag\\
    & +\frac{|\m_2|K_1(2|\m_2\sin\tfrac \phi 2|)
      -|\m_1|K_1(2|\m_1\sin\tfrac \phi 2|)}{(\m_2^2-\m_1^2)^2|\sin \tfrac \phi 2|}\biggr]
      \frac{Z(\vec\m,\theta-\phi)}{Z(\vec\m,\theta)}
      +\biggl[|\sin\tfrac{\phi}{2}|\sign(u_1)\frac{\m_1\cos\phi-\m_2\cos(\theta-\phi)}
      {\m_2^2-\m_1^2} K_1(2|\m_1\sin\tfrac{\phi}{2}|)\notag\\
    & +(\m_1\m_2\cos(\theta -\phi)-\m_2^2\cos\phi)
      \frac{K_0(2|\m_1\sin\tfrac{\phi}{2}|)  -K_0(2|\m_2\sin\tfrac{\phi}{2}|)}
      {(\m_2^2-\m_1^2)^2}\biggr]
      \frac{2\zeta_0(\vec\m,\theta-\phi)}{Z(\vec \m,\theta)}\biggl\}\,.
    \label{chiralcond-nzm-2f.b}
  \end{align}
\end{widetext}
The limit $\m_2\to\infty$ yields the one-flavor
result~\eqref{sigma-dyn-therm-f1}, i.e.,
$\Sigma_{1,N_f=2}^\d(\m_1,\m_2=\infty,\theta)=\Sigma_{N_f=1}^\d(\m_1,\theta)$. This can be
readily checked because only two terms of the integral are of leading
order and the Bessel function $I_\nu$ can be approximated as in
Eq.~\eqref{eq:Iasympt}. In the limit $\m_1\to\infty$ (at fixed $\m_2$) one
can show that $\Sigma_1^\d$ is proportional to $1/\m_1$.

\begin{figure}[b]
  \centerline{\includegraphics[width=\columnwidth]{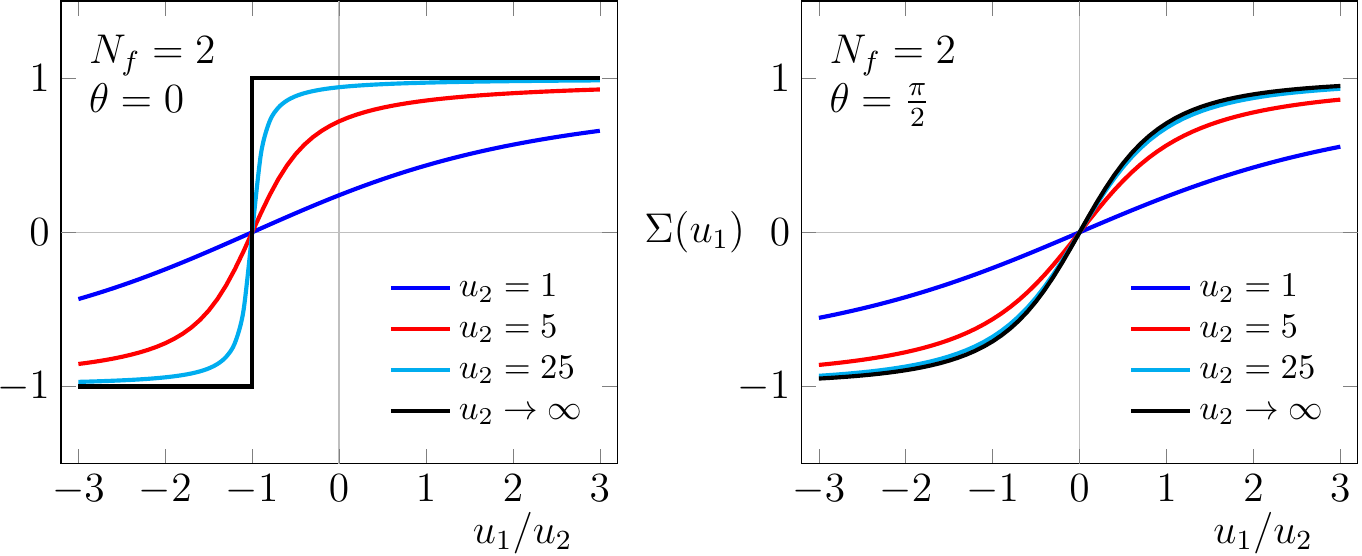}}
  \caption{Chiral condensate of the first quark in the two-flavor
    theory, see Eq.~\eqref{chi-cond-Nf2-mass-theta-m1}, as a function
    of the first quark mass $\m_1$ rescaled w.r.t.\ the fixed second
    quark mass $\m_2$. The black curve is the thermodynamic limit. The
    jump in the left plot ($\theta=0$) is the Dashen
    point~\cite{Dashen:1970et,Creutz:2013xfa}, which corresponds to a
    first-order phase transition.}
  \label{fig:chiralNf2a}
\end{figure}

We emphasize that $\Sigma_1^\d$ remains finite for large masses. The reason
is the same as discussed in section \ref{section-4}, namely that the
exponents of the $K_\nu$ and of the partition function cancel.  Of
course this should happen because the total chiral condensate, which
can be obtained from the mass derivative~\eqref{eq:Sigmam} of the
two-flavor partition function,
\begin{align}
  \label{chi-cond-Nf2-mass-theta-m1}
  &\frac{\Sigma_1(\M,\theta)}\Sigma=\frac d{d\m_1}\log Z(\M,\theta)\\
  &=\frac{\m_1+\m_2\cos\theta}{\sqrt{\m_1^2+\m_2^2+2\m_1\m_2\cos\theta}}
    \frac{I_2(\sqrt{\m_1^2+\m_2^2+2\m_1\m_2\cos\theta}) }{I_1(\sqrt{\m_1^2+\m_2^2+2\m_1\m_2\cos\theta}) }\,,\notag
\end{align}
is finite.  Although the expression for $\Sigma_1$ is quite complicated
when derived via Eqs.~\eqref{sigsum0} and~\eqref{chiralcond-nzm-2f.b},
we checked numerically that it agrees with
Eq.~\eqref{chi-cond-Nf2-mass-theta-m1}. (We did not succeed to give a
direct analytical proof.)  The behavior of $\Sigma_1$ is illustrated in
Fig.~\ref{fig:chiralNf2a} for several masses at the two angles
$\theta=0$ and $\pi/2$.  It changes sign at $\m_1/\m_2=-\cos\theta$, which becomes
the Dashen point~\cite{Dashen:1970et,Creutz:2013xfa} for
$\theta=0,\pi$ when taking the thermodynamic limit.
 
\begin{figure}[t!]
  \centerline{\includegraphics[width=.9\columnwidth]{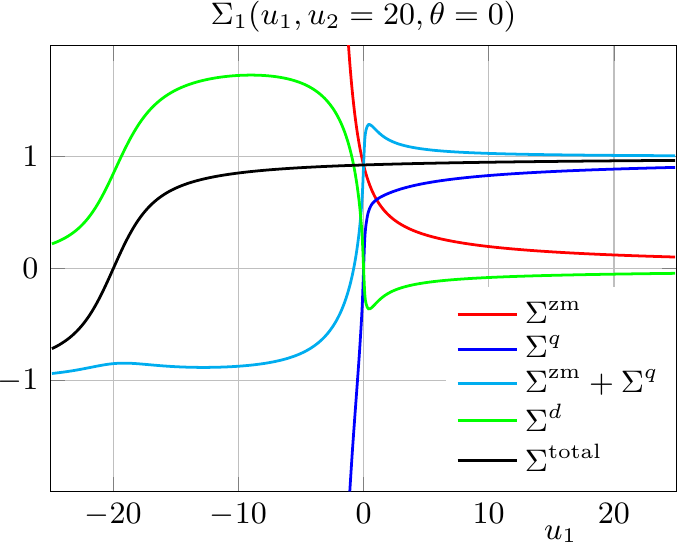}}
  \caption{Mass dependence of the various contributions to the chiral
    condensate for two-flavor QCD at $\theta = 0$. The chiral condensate of
    the first quark is shown as a function of $\m_1=m_1 V \Sigma$, while
    the second mass is kept fixed at $\m_2=20$.}
  \label{fig:sigdec20}
\end{figure}

In Fig.~\ref{fig:sigdec20} we show the decomposition of the chiral
condensate for $\theta=0$ with one mass kept fixed at $\m_2 =20$. The sum
of the quenched part and the zero-mode part, $\Sigma^\q+\Sigma^{\zm}$, results
in a chiral condensate with a discontinuity in the thermodynamic limit
at $\m_1=0$.  Both parts become exponentially large in the volume when
the product of the quark masses is negative, but their sum is finite.
The dynamical part of the spectral density results in a chiral
condensate with a discontinuity at $\m_1=0$ that cancels the
discontinuity of $\Sigma^\q+\Sigma^{\zm}$, and creates a new discontinuity at
$\m_1 = -\m_2$.

\subsection{Thermodynamic limit}\label{sec:Nf2-limit}

Before we turn to the general setting we first consider the cases
$\theta = 0$ with $\m_1\m_2>0$ and $\theta = \pi$ with
$ \m_1 \m_2 < 0$, in which there is no sign problem.  In the
thermodynamic limit, the chiral condensate becomes
\begin{equation}
  \label{chi-cond-Nf2-mass-theta-m1-limit-special}
  \frac{\Sigma_1(\M,\theta)}\Sigma=\sign(\m_1)\frac{I_2(|\m_1|+|\m_2|)}{I_1(|\m_1|+|\m_2|)}\overset{|\m_1|,|\m_2|\gg1}{\approx}\sign(\m_1)
\end{equation}
because the leading asymptotic behavior of the Bessel functions does
not depend on the index.  Hence, its behavior is not different from
the one-flavor case. The difference between the one- and two-flavor
theory shows up in the zero-mode contribution, which is
\begin{equation}\label{chiralcond-zero-Nf2-theta-limit-special}
  \frac{\Sigma^{\zm}_1(\M,\theta)}\Sigma\overset{|\m_1|,|\m_2|\gg1}{\approx}
  \sign(\m_1)\sqrt{\frac{2|\m_2|}{\pi|\m_1|(|\m_1|+|\m_2|)}}\,.
\end{equation}
To obtain this result for $\theta=0$ and $\m_1\m_2>0$ we have performed a
saddle-point expansion of the
integrand~\eqref{chiralcond-zero-Nf2-theta-special} about the point
$\phi=0$. Hence, we again have an algebraic dependence on the quark
masses, which has a similar behavior as in the one-flavor case when
both masses are equal,
$\Sigma_1^{\zm}(|\m_1|=|\m_2|=\m,0)/\Sigma\approx1/\sqrt{\pi \m}$, cf.\
Eq.~\eqref{density-zero-Nf1-theta-special-limit}.

The situation changes drastically when there is a sign problem, i.e.,
$\cos\theta\neq \sign(\m_1\m_2)$.  In this case the zero-mode contribution
again exhibits exponential behavior,
\begin{align}
  &\frac{\Sigma_1^{\zm}(\M,\theta)}\Sigma\approx \frac{1}{\sqrt{8\pi}\sin^2\left[\frac\theta2+(1-\sign(\m_1\m_2))\frac\pi 4\right]}\frac{1}{u_1}\notag\\
  &\times\frac{(\m_1^2\!+\!\m_2^2\!+\!2\m_1\m_2\cos\theta)^{3/2}}{\sqrt{|\m_1\m_2|}(|\m_1|+|\m_2|)}
    e^{|\m_1| +|\m_2|-\sqrt{\m_1^2+\m_2^2+2\m_1\m_2\cos\theta}}\,.
  \label{chiralcond-zero-Nf2-theta-limit}
\end{align}
To obtain this result we performed a saddle-point approximation of
Eq.~\eqref{eq:sigmazm} about the point
$\varphi=\theta+(1-\sign(\m_1\m_2))\pi/2$. Note that the derivation of
\eqref{chiralcond-zero-Nf2-theta-limit} is only valid for both
$|u_1| \gg 1$ and $|u_2| \gg 1$ and cannot be used in the chiral limit of
any of the quark masses.  For $\sign(\m_1\m_2)>0$ and $\theta\to\pi$, the
exponential divergence in Eq.~\eqref{chiralcond-zero-Nf2-theta-limit}
is given by $\exp(2\min\{|\m_1|,|\m_2|\})$, which changes drastically
at the point $|\m_1|=|\m_2|$ from $\exp(2|\m_1|)$ to $\exp(2|\m_2|)$.

\begin{figure}[t!]
  \centerline{\includegraphics[width=0.9\columnwidth]{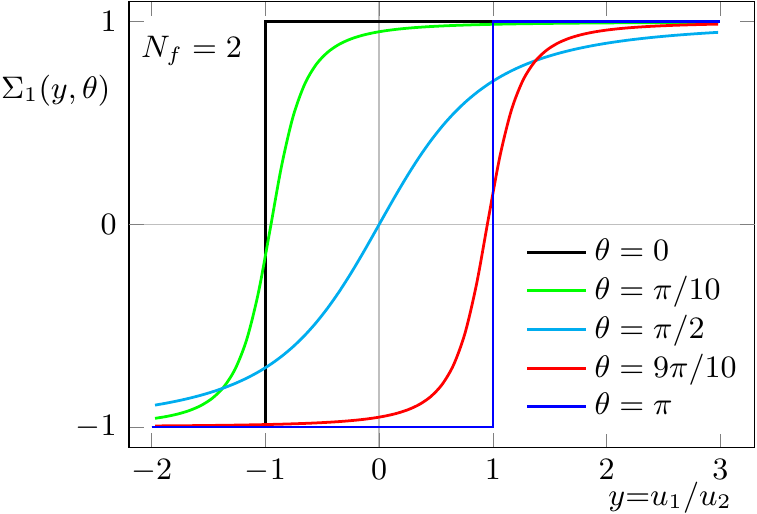}}
  \caption{Thermodynamic limit of the chiral
    condensate~\eqref{chi-cond-Nf2-limit} as a function of the first
    quark mass $\m_1$ for several $\theta$-angles and $\m_2>0$. The chiral
    condensate only depends on the ratio $\m_1/\m_2$ in this
    particular limit. The Dashen point only exists for $\theta=0,\pi$ and
    shows up as a jump in the chiral condensate, reflecting the nature
    of a first-order phase transition.}
  \label{fig:chiralNf2c}
\end{figure}

The total chiral condensate still behaves algebraically in the masses,
\begin{align}\label{chi-cond-Nf2-limit}
  &\frac{\Sigma_1(\M,\theta)}\Sigma\overset{|\m_1|,|\m_2|\gg1}{\approx}
    \frac{\m_1+\m_2\cos\theta}{\sqrt{\m_1^2+\m_2^2+2\m_1\m_2\cos\theta}}\notag\\
  &=\sign(\m_2)\frac{y+\cos\theta}{\sqrt{y^2+1+2y\cos\theta}}\quad\text{with}\quad y=\m_1/\m_2\notag\\
  &=
    \begin{cases}
      \sign(\m_1+\m_2)\,, & \theta=0\,,\\
      \sign(\m_1-\m_2)\,, & \theta=\pi\,.
    \end{cases}
\end{align}
Apart from the factor $\sign(\m_2)$, the thermodynamic limit of the
chiral condensate is a function of $\m_1/\m_2$ only and is shown in
Fig.~\ref{fig:chiralNf2c} for several values of $\theta$. It changes sign
at $\m_1/\m_2=-\cos\theta$. For $\theta\neq0,\pi$ this transition is smooth, but for
$\theta=0$ or $\theta=\pi$ (corresponding to the Dashen
point~\cite{Dashen:1970et,Creutz:2013xfa}) it is discontinuous.

\section{Conclusions}\label{sec:conclusio}

At nonzero $\theta$-angle, the discontinuity of the chiral condensate does
in general not coincide with the support of the Dirac spectrum. In
particular, for one flavor QCD there is no discontinuity at zero quark
mass, and for two flavor QCD with quark masses $m_1$ and $m_2$, the
chiral condensate of the first quark does not have a discontinuity at
$m_1 =0$ but rather at $m_1 = -m_2$. We have analyzed this behavior in
terms of the contribution from the zero modes, the contribution of the
mass independent part of the Dirac spectrum (the ``quenched'' part),
and the contribution of the remainder of the Dirac spectrum which is
sensitive to the fermion determinant (at fixed topological charge).

At fixed $\theta$-angle, we have obtained a compact general formula for the
contribution of the zero modes and of the quenched part of the Dirac
spectrum to the chiral condensate.  Both formulas are valid for any
number of flavors and are given by an integral over a flavor
independent kernel times the ratio of $N_f$ flavor partition
functions. The formula for the zero modes is completely general, while
the expression for the quenched part has been obtained in the
microscopic domain of QCD but is also valid for any number of flavors.
Both contributions diverge exponentially with the volume at nonzero
$\theta$-angle, but the divergent contributions cancel identically when
added, leaving a result that is finite in the thermodynamic limit.
The deeper reason for the cancellation based on general ideas from
spectral flow and topology, is that when we have $|\nu|$ zero modes, the
spectrum near zero is depleted by $|\nu|$ modes, half of them with
positive eigenvalues and the other half with negative
eigenvalues. This depletion gives a contribution to the chiral
condensate with the opposite sign. For large $\m=mV\Sigma$ it does not
matter whether the modes are exactly at zero or are close to zero.
The fermion determinant results in an additional eigenvalue repulsion
from zero, which does not depend on $\nu$, and the contribution to the
chiral condensate due to this modification of the Dirac spectrum is
expected to remain finite in the thermodynamical limit.

For one and two flavors, we have also obtained exact analytical
expressions for the dynamical part of the Dirac spectrum and the
chiral condensate in the microscopic or $\epsilon$ domain of QCD which
confirm the above picture.

For $\theta \ne 0$ both the quenched and dynamical contribution to the
spectral density as well as their sum are strongly oscillating with an
amplitude that diverges exponentially with the volume and a period on
the order of $1/V$.  From QCD at nonzero chemical potential we have
learnt that this behavior may cancel the discontinuity of the chiral
condensate and shift it to a different point.  The sum of the quenched
part and the zero mode part retain the discontinuity at $m_1 =0$, but
the oscillations of the dynamical part cancels this discontinuity and
move it to $m_1 = -m_2$ for two flavors or to infinity for one flavor.
The effect of the zero modes is to create a gap at zero but, in the
quenched approximation, the position of the remaining eigenvalues does
not depend very much on $\nu$.  Since the condensate is obtained in the
thermodynamical limit, it is not surprising that the quenched
contributions to the chiral condensate have a discontinuity at
$m_1= 0$. We thus conclude that the determinant introduces
correlations in the Dirac spectrum that cancel the discontinuity at
$m_1 =0$ and move it to $m_1 = -m_2$. Currently we do not have a good
understanding of the nature of these correlations, but hope to return
to this issue in future work.

We have also seen that the correct computation of the chiral
condensate at nonzero $\theta$-angle requires a subtle balance between zero
and nonzero modes. Even the slightest incompatibilities will give
results that are completely off. This will make lattice QCD
simulations at nonzero $\theta$-angle a formidable, and perhaps impossible,
task, and we have to rely on analytical work to make further progress.

\section*{Acknowledgments}

MK acknowledges support by the German Research Foundation (DFG) via
CRC 1283: ``Taming uncertainty and profiting from randomness and low
regularity in analysis, stochastics and their applications". TW was
supported by the DFG in the framework of SFB/TRR-55.  JV was partially
supported by U.S.\ DOE Grant No.\ DE-FAG-88FR40388.

\appendix

\section{Resummation of products of Bessel functions}\label{sec:resum}

To compute the sums of Bessel functions needed in the main text we
consider the sum
\begin{align}\label{a1.1}
  \Xi_{a}^{(N)}(\theta,\M)=\sum_{\nu=-\infty}^\infty e^{i\nu\theta}\prod\limits_{j=1}^N I_{\nu+a_j}(\m_j)\,,
\end{align}
where $I_\nu$ is the modified Bessel function of the first kind. The
indices $a=(a_1,\ldots,a_{N})$ are chosen to be integers while the masses
$\M=(\m_1,\ldots,\m_N)$ can be arbitrary, even complex valued. Indeed we
need imaginary masses $\M$ to generate the Bessel functions of the
first kind via the relation $I_\nu(iz)=i^\nu J_\nu(z)$.

Using an integral representation of the Bessel function we can sum
over $\nu$ employing the relation
$\sum_{\nu=-\infty}^\infty e^{i\nu\phi} = 2\pi\delta(\phi)$. Note that the argument of the Dirac
delta function has to be taken modulo $2\pi$, which we omit. This sum
yields
\begin{align}\label{a1.2}
  \Xi_{a}^{(N)}(\theta,\M)=2\pi\prod_{j=1}^{N}\int_{-\pi}^\pi\!\frac{d\phi_j}{2\pi}\,
  e^{ia_j\phi_j+\m_j\cos\phi_j}\delta\Bigl(\sum_{k=1}^N\phi_k\!+\!\theta\Bigr).
\end{align}
The simplest case is $N=1$, for which
\begin{align}
 \Xi_{a_1}^{(1)}(\theta,\m_1)=e^{-ia_1\theta}e^{\m_{1}\cos\theta}\,.
\end{align}
For $a=0$ this corresponds to the one-flavor partition function
\begin{equation}
  Z(\m_1,\theta)=\Xi_{0}^{(1)}(\theta,\m_1)=e^{\m_{1}\cos\theta}\,.\label{a1.8}
\end{equation}
For $N>1$, we define new variables $\vartheta_1=\phi_1$ and
$\vartheta_j=\phi_j+\vartheta_{j-1}$ for $j=2,\ldots,N$. This allows us to
evaluate the delta function, and we end up with a chain of integrals,
\begin{align}
 \label{a1.3}
  \Xi_{a}^{(N)}(\theta&,\M)
  =e^{-ia_N\theta}\prod_{j=1}^{N-1}\int_{-\pi}^\pi\frac{d\vartheta_j}{2\pi}\,
    e^{i(a_j-a_{j+1})\vartheta_j}\\
  &\times e^{\m_1\cos\vartheta_1+\m_N\cos(\theta+\vartheta_{N-1})+\sum_{k=2}^{N-1}\m_k\cos(\vartheta_k-\vartheta_{k-1})}
    \,.\notag
\end{align}
For general even $N$ we set $N=2n$ with $n\in\mathbb{N}$ and integrate
over $\vartheta_1,\vartheta_3,\ldots,\vartheta_{2n-1}$. This again produces
Bessel functions,
\begin{widetext}
\begin{align}
  &\Xi_{a}^{(2n)}(\theta,\M)=e^{-ia_{2n}\theta}\prod_{j=1}^{n-1}
    \int_{-\pi}^\pi\frac{d\vartheta_{2j}}{2\pi}\,
    e^{i(a_{2j}-a_{2j+1})\vartheta_{2j}}
    \left(\frac{\m_1+\m_2e^{i\vartheta_2}}{\m_1+\m_2e^{-i\vartheta_2}}\right)^{(a_1-a_2)/2}
    I_{a_1-a_2}\Bigl(\sqrt{\m_1^2+\m_2^2+2\m_1\m_2\cos\vartheta_2}\Bigr)\notag\\
  &\times \left(\frac{\m_{2n-1}e^{i\vartheta_{2n-2}}+\m_{2n}e^{-i\theta}}
    {\m_{2n-1}e^{-i\vartheta_{2n-2}}+\m_{2n}e^{i\theta}}\right)^{(a_{2n-1}-a_{2n})/2}
    I_{a_{2n-1}-a_{2n}}\Bigl(\sqrt{\m_{2n-1}^2+\m_{2n}^2+2\m_{2n-1}\m_{2n}\cos(\vartheta_{2n-2}+\theta)}\Bigr)
    \label{a1.4}\\
  &\times\prod\limits_{k=1}^{n-2}\left(\frac{\m_{2k+1}e^{i\vartheta_{2k}}+\m_{2k+2}e^{i\vartheta_{2k+2}}}
    {\m_{2k+1}e^{-i\vartheta_{2k}}+\m_{2k+2}e^{-i\vartheta_{2k+2}}}\right)^{(a_{2k+1}-a_{2k+2})/2}\!
    I_{a_{2k+1}-a_{2k+2}}\left(\sqrt{\m_{2k+1}^2\!+\!\m_{2k+2}^2\!+\!2\m_{2k+1}\m_{2k+2}
    \cos(\vartheta_{2k}\!-\!\vartheta_{2k+2})}\right).\notag
\end{align}
In the case of $N=2$, which is needed for the partition function of
two flavors and can also be used for the quenched level density, we
obtain
\begin{align}
  \Xi_{a_1,a_2}^{(2)}(\theta,\m_1,\m_2)
  =e^{-i(a_1+a_{2})\theta/2}
  \left(\frac{\m_{1}e^{i\theta}+\m_{2}}{\m_{2}e^{i\theta}+\m_{1}}\right)^{(a_{1}-a_{2})/2}
  I_{a_{1}-a_{2}}\Bigl(\sqrt{\m_{1}^2+\m_{2}^2+2\m_{1}\m_{2}\cos\theta}\Bigr)\,.
  \label{a1.5}
\end{align}
Then the two-flavor partition function is
\begin{align}
  Z(\m_1,\m_2,\theta)
  =2\frac{\m_2\Xi_{0,1}^{(2)}(\theta,\m_1,\m_2)-\m_1\Xi_{1,0}^{(2)}(\theta,\m_1,\m_2)}{\m_2^2-\m_1^2}
  =\frac{2I_{1}\bigl(\sqrt{\m_{1}^2+\m_{2}^2+2\m_{1}\m_{2}\cos\theta}\bigr)}
    {\sqrt{\m_{1}^2+\m_{2}^2+2\m_{1}\m_{2}\cos\theta}}\,.
    \label{a1.5b}
\end{align}
For $N=4$, which is needed for the level density with two flavors, the
result is a single integral,
\begin{align}
  \Xi_{a}^{(4)}(\theta,\m_1,\m_2,\m_3,\m_4)&=e^{-ia_{4}\theta}\int_{-\pi}^\pi\frac{d\phi}{2\pi}\,
    e^{i(a_{2}-a_{3})\phi}\left(\frac{\m_1+\m_2e^{i\phi}}{\m_1+\m_2e^{-i\phi}}\right)^{(a_1-a_2)/2} 
    \left(\frac{\m_{3}e^{i\phi}+\m_{4}e^{-i\theta}}{\m_{3}e^{-i\phi}+\m_{4}e^{i\theta}}\right)^{(a_{3}-a_{4})/2}
                                             \notag\\
  &\quad\times I_{a_1-a_2}\Bigl(\sqrt{\m_1^2+\m_2^2+2\m_1\m_2\cos\phi}\Bigr)
    I_{a_{3}-a_{4}}\Bigl(\sqrt{\m_{3}^2+\m_{4}^2+2\m_{3}\m_{4}\cos(\phi+\theta)}\Bigr)\,.
    \label{a1.6}
\end{align}
For general odd $N$ we set $N=2n+1$ with $n\in\mathbb{N}$ and again
integrate over $\vartheta_1,\vartheta_3,\ldots,\vartheta_{2n-1}$. This
leads to a slightly different result,
\begin{align}
  &\Xi_{a}^{(2n+1)}(\theta,\M)=e^{-ia_{2n+1}\theta} \notag\\
  \label{a1.7}
  &\times \prod_{j=1}^{n}\int_{-\pi}^\pi\frac{d\vartheta_{2j}}{2\pi}\,
    e^{i(a_{2j}-a_{2j+1})\vartheta_{2j}} e^{\m_{2n+1}\cos(\theta+\vartheta_{2n})}
    \left(\frac{\m_1+\m_2e^{i\vartheta_2}}{\m_1+\m_2e^{-i\vartheta_2}}\right)^{(a_1-a_2)/2}
    I_{a_1-a_2}\Bigl(\sqrt{\m_1^2+\m_2^2+2\m_1\m_2\cos\vartheta_2}\Bigr) \\
  &\times\prod\limits_{k=1}^{n-1}
    \left(\frac{\m_{2k+1}e^{i\vartheta_{2k}}+\m_{2k+2}e^{i\vartheta_{2k+2}}}
    {\m_{2k+1}e^{-i\vartheta_{2k}}+\m_{2k+2}e^{-i\vartheta_{2k+2}}}\right)^{(a_{2k+1}-a_{2k+2})/2}
    I_{a_{2k+1}-a_{2k+2}}\left(\sqrt{\m_{2k+1}^2\!+\!\m_{2k+2}^2\!+\!2\m_{2k+1}\m_{2k+2}
    \cos(\vartheta_{2k}\!-\!\vartheta_{2k+2})}\right). \notag
\end{align}
For $N=3$, which is employed for the level density with one flavor,
the result is a single integral,
\begin{align}
  \Xi_{a}^{(3)}(\theta,\m_1,\m_2,\m_3)
  =e^{-ia_{3}\theta}\int_{-\pi}^\pi\frac{d\phi}{2\pi}\,
  e^{i(a_{2}-a_{3})\phi}e^{\m_{3}\cos(\theta+\phi)}
  \left(\frac{\m_1+\m_2e^{i\phi}}{\m_1+\m_2e^{-i\phi}}\right)^{(a_1-a_2)/2}
  I_{a_1-a_2}\Bigl(\sqrt{\m_1^2+\m_2^2+2\m_1\m_2\cos\phi}\Bigr)\,.\label{a1.9}
\end{align}
The results above simplify further for the specific sums we are
considering. For the contribution of the nonzero modes to the
two-flavor level density we need $N=3$ and $4$ with $\m_1=\m_2=ix$ and
$\theta\to\theta-\pi$, i.e.,
\begin{align}
  \label{a1.10}
  \sum_{\nu=-\infty}^\infty e^{i\nu\theta}J_{\nu+a_1}(x)J_{\nu+a_2}(x)I_{\nu+a_3}(\m)
  &=e^{-i\pi (a_1+a_2)/2}\Xi_{a}^{(3)}(\theta-\pi,ix,ix,\m)\\
  &=(-1)^{a_2-a_3}e^{-ia_{3}\theta}\int_{-\pi}^\pi\frac{d\phi}{2\pi}\,
    e^{i(a_1+a_2-2a_{3})\phi/2}e^{-\m\cos(\theta+\phi)} J_{a_1-a_2}\left(2x\cos\tfrac{\phi}{2}\right)\notag
\end{align}
and
\begin{align}
  &\sum_{\nu=-\infty}^\infty e^{i\nu\theta}J_{\nu+a_1}(x)J_{\nu+a_2}(x)I_{\nu+a_3}(\m_1)I_{\nu+a_4}(\m_2)
  =e^{-i\pi (a_1+a_2)/2}\Xi_{a}^{(4)}(\theta-\pi,ix,ix,\m_1,\m_2)\notag\\
  &=(-1)^{a_2+a_4}e^{-i(a_3+a_{4})\theta/2}\int_{-\pi}^\pi\frac{d\phi}{2\pi}\,
    e^{i(a_1+a_{2}-a_{3}-a_4)\phi/2}
    \left(\frac{\m_1e^{i(\theta+\phi)}-\m_2}{\m_1-\m_2e^{i(\theta+\phi)}}\right)^{(a_{3}-a_{4})/2}\notag\\
  &\qquad\qquad\qquad\qquad\qquad\qquad\qquad\quad\times
    J_{a_1-a_2}\left(2x\cos\tfrac{\phi}{2}\right)
    I_{a_{3}-a_{4}}\Bigl(\sqrt{\m_1^2+\m_2^2-2\m_1\m_2\cos(\theta+\phi)}\Bigr)\,.
    \label{a1.11}
\end{align}

\section{\boldmath Calculation of the condensate $\Sigma_1^\d$ for two
  flavors}\label{sec:4}

The dynamical part of the level density is given by
Eq.~\eqref{level-quenched-f2}. We combine this result with
Eq.~\eqref{eq:sigma}. Thus we have to evaluate the integral
\begin{align}
  \Sigma_1^\d(\M,\theta)
  &=\int_0^\infty d\x\,\frac{2\m_1}{\x^2+\m_1^2} \rho^\d(\x,\M,\theta)\notag\\
  &=-\frac{2\m_1}{Z(\M,\theta)}\int_0^\infty d\x\int_{-\pi}^\pi\frac{d\phi}{2\pi}\,
    \frac{\x}{(\x^2+\m_1^2)^2(\x^2+\m_2^2)}
    \biggl\{\x \bigl[2\m_1\m_2\sin\tfrac{\theta- \varphi}2 +(\m_2^2+\m_1^2)\sin\tfrac\varphi2\bigr]
    J_1(2\x\sin\tfrac \phi 2)Z(\M,\theta-\phi)\notag\\ 
  &\quad +2\left[\x^2\cos\varphi+\m_1\m_2\cos(\theta-\varphi)\right] J_0(2\x\sin\tfrac \phi 2)
    I_0\Bigl(\sqrt{\m_1^2+\m_2^2 +2\m_1\m_2\cos(\theta-\varphi)} \Bigr)\biggl\}\,.
\end{align}
As the first step we perform a partial fractions expansion of the
ratios
\begin{align}
  \frac{1}{(\x^2+\m_1^2)^2(\x^2+\m_2^2)} &=\frac{1}{\m_2^2-\m_1^2}\left[\frac{1}{(\x^2+\m_1^2)^2}-\frac{1}{\m_2^2-\m_1^2}\frac{1}{\x^2+\m_1^2}+\frac{1}{\m_2^2-\m_1^2}\frac{1}{x^2+\m_2^2}\right],\notag\\
 \frac{\x^2\cos\varphi+\m_1\m_2\cos(\theta-\varphi)}{(\x^2+\m_1^2)^2(\x^2+\m_2^2)} &=\frac{\m_1\m_2\cos(\theta-\varphi)-\m_1^2\cos\varphi}{(\m_2^2-\m_1^2)(\x^2+\m_1^2)^2}+\frac{\m_2^2\cos\varphi-\m_1\m_2\cos(\theta-\varphi)}{(\m_2^2-\m_1^2)^2(\x^2+\m_1^2)}-\frac{\m_2^2\cos\varphi-\m_1\m_2\cos(\theta-\varphi)}{(\m_2^2-\m_1^2)^2(x^2+\m_2^2)}
\end{align}
for the first and second term in the integral, respectively. For the
integral over the first term we need the integrals~\eqref{eq:J13}
and~\eqref{eq:J12} while for the second term we employ~\eqref{eq:J02}
and~\eqref{eq:J01}. After some algebra we find
Eq.~\eqref{chiralcond-nzm-2f.b}.

\end{widetext}
\section{Integrals over Bessel functions}
\label{app:integrals}

At several places of our work we need recurrence relations and other
identities of Bessel functions which can be found in
\cite{mathematica,Gradshteyn:2007}. We will briefly summarize those we
need here.  The ordinary and modified Bessel functions of the first
kind satisfy the recurrence relations
\begin{align}
  \x(J_{\nu+1}(\x)+J_{\nu-1}(\x))&=2\nu J_{\nu}(\x)\,,\label{J.rec}\\
  \m(I_{\nu-1}(\m)-I_{\nu+1}(\m))&=2\nu I_\nu(\m)\,,\label{I.rec}\\
\partial_y \frac{I_\nu(\sqrt{y})}{y^{\nu/2}}&=\frac{1}{2}\frac{I_{\nu+1}(\sqrt{y})}{y^{(\nu+1)/2}}\,.\label{I.rec.diff}
\end{align}
The modified Bessel function of the second kind also satisfies two
recursion relations~\cite{Gradshteyn:2007},
\begin{align}
\x(K_{\nu+1}(\x)-K_{\nu-1}(x))&=2\nu K_{\nu}(\x)\,,\label{Bessel-K-rec.a}\\
K_{\nu+1}(\x)+K_{\nu-1}(\x)&=-2\partial_{\x}K_\nu(\x)\,.\label{Bessel-K-rec.b}
\end{align}
Specifically, we have
\begin{align}
\partial_{\x}K_{0}(a\x)&=-a K_{1}(a\x)\,,\label{Bessel-K-dif.a}\\
\partial_{\x}[\x K_{1}(a\x)]&=-ax K_{0}(a\x)\,.\label{Bessel-K-dif.b}
\end{align}
Moreover we need the integral identities
\begin{align}
  \label{eq:J11}
  \int_0^\infty d\x\, \frac{J_1(2\x t) }{\x^2+\m^2} 
  &= \frac1{2t\m^2}-\frac{\sign(t)}{|\m|}K_1(2|t\m|)\,,\\
  \label{eq:J13}
  \int_0^\infty d\x\, \frac{\x^2J_1(2\x t)}{\x^2+\m^2} &= \sign(t) |\m|K_1(2|t\m|)\,,\\
  \label{eq:J12}
  \int_0^\infty d\x\, \frac{\x^2J_1(2\x t)}{(\x^2+\m^2)^2} &= tK_0(2|t\m|)\,,\\
  \label{eq:J02}
  \int_0^\infty d\x\, \frac{\x J_0(2\x t)}{\x^2+\m^2} &= K_0(2|t\m|)\,,\\
  \label{eq:J01}
  \int_0^\infty d\x\, \frac{\x J_0(2\x t)}{(\x^2+\m^2)^2} &= \frac{|t|}{|\m|}K_1(2|t\m|)\,.
\end{align}
The first, second and fourth integral were also given in
\cite{Verbaarschot:2014upa}, where it was implicitly assumed that
$t>0$.

\section{\boldmath Thermodynamic limit of $\Sigma^\d$ for $N_f=1$}
\label{app:asympt}

In this appendix we derive Eq.~\eqref{sig-d-f1-lim}.  We consider the
RHS of \eqref{sigma-dyn-therm-f1}, which we denote by $I$.  For large
argument the Bessel functions $K_\nu(x)$ can be approximated by
$e^{-x}\sqrt{\pi/2x}$. Therefore, in a saddle-point approximation, the
exponent to be analyzed is
\begin{align}
  f(\phi)&=\m\cos(\theta-\phi)-\m\cos\theta-2|\m\sin\tfrac\phi2|\notag\\
  &=-2|u\sin\tfrac\phi2|\left[1-\sign(u\sin\tfrac\phi2)\sin(\theta-\tfrac\phi2)\right],
\end{align}
which is always non-positive and has a maximum of
$f_{\text{max}}(\phi)=0$. A straightforward analysis shows that for
$\m\cos\theta>0$ the maximum is assumed only at $\phi=0$, while for
$\m\cos\theta<0$ it is also assumed at $\bar\phi=2\theta-(2k+1)\pi$, where
$k\in\mathbb Z$ has to be chosen such that $\bar\phi\in[-\pi,\pi]$. The latter is
a true saddle point and dominates the integral. Expansion about
$\bar\phi$ yields to leading order in $|\m|$
\begin{align}
  I\approx2\cos\theta\qquad(\text{if }\m\cos\theta<0)\,.
\end{align}

For $\m\cos\theta>0$ we have to expand about $\phi=0$. This is not a true
saddle point since the derivative of $f(\phi)$ is nonzero and
discontinuous at this point. Furthermore, for $\phi\to0$ we cannot use the
asymptotic expansion of $K_\nu(x)$. Since the term involving $K_0$ comes
with an additional factor of $\sin\frac\phi2$ it is subleading and can be
dropped. Hence, to leading order in $|\m|$,
\begin{align}
  I&\approx-2\int_{-\pi}^\pi \frac{d\phi}{2\pi}\,\frac{e^{\m\cos(\theta-\phi)}}{e^{\m\cos\theta}}
     |\m\sin\tfrac\phi2|\cos\theta K_1(2|\m\sin\tfrac\phi2|)\notag\\
   &\approx-\frac{\cos\theta}{2\pi|\m|}\int_{-\infty}^\infty dt\,e^{t\sin\theta\sign(u)}|t|K_1(|t|)\notag\\
   &=-\frac{\cos\theta}{\pi|\m|}\int_0^\infty dt\,\cosh(t\sin\theta)tK_1(t)\notag\\
   &=-\frac{\cos\theta}{2|u\cos^3\theta|}
     =-\frac1{2\m\cos^2\theta}\quad(\text{if }\m\cos\theta>0)\,,
\end{align}
where the second line was obtained by transforming $\phi=t/|\m|$ and
expanding in $t$. The integral in the third line equals
$\pi/2(1-\sin^2\theta)^{3/2}=\pi/2|\cos\theta|^3$.

Finally we consider the case of $\cos\theta=0$. It is straightforward to
show that the same result for $I$ is obtained for $\theta=\pm\pi/2$ and that
the result is odd in $\m$. Hence we only consider $\theta=\pi/2$ and
$\m>0$ in the following. For $\theta\to\pi/2$ we have $\bar\phi\to0$, i.e., we again
have to expand about $\phi=0$. In this case we find for the exponent to
leading order in $\phi$
\begin{align}
  f(\phi)\approx\begin{cases}
         -\frac18u\phi^3\,, & \phi>0\,,\\
         2u\phi\,,  & \phi<0\,.
        \end{cases}
\end{align}
The dominant contribution to the integral is thus obtained from the
region $\phi>0$. Since we are expanding for small $\phi$, the term involving
$K_1$ gives twice the result of the term involving $K_0$. Using the
asymptotic expansion of $K_\nu$ we obtain to leading order in $|\m|$
\begin{align}
  I&\approx-\frac{3u}\pi\int_0^\pi d\phi\,e^{-\frac18u\phi^3}\frac{\phi^2}4\frac{\sqrt\pi}{\sqrt{2u\phi}}\notag\\
   &\approx-\frac3{\sqrt\pi u^{1/3}}\int_0^\infty dt\,e^{-t^3}t^{3/2}\notag\\
   &=-\frac{\Gamma(5/6)}{\sqrt\pi u^{1/3}}\qquad(\text{if }\cos\theta=0\text{ and }\m>0)\,.
\end{align}
Observing that $I$ is odd in $\m$ for $\cos\theta=0$ we obtain the last
line of Eq.~\eqref{sig-d-f1-lim}.

\bibliography{KVW}

\end{document}